\documentclass[11pt]{article}
\pdfoutput=1
\usepackage[utf8]{inputenc}
\usepackage{graphicx}
\usepackage{amssymb}
\usepackage{amsmath}
\usepackage{subfigure}
\usepackage{booktabs}
\usepackage{multirow}
\usepackage[title]{appendix}
\usepackage{dcolumn}
\newcolumntype{d}[1]{D{.}{.}{#1}}

\usepackage{soul,color} 
\sethlcolor{yellow} 

\def\dec{\pi^0 \to \mathrm{invisible}}
\def\kppg{K^+\to\pi^+\pi^0(\gamma)}
\def\kppgIB{K^+\to\pi^+\pi^0(\gamma_\mathrm{IB})}

\def\kmng{K^+ \to \mu^+ \nu_{\mu} (\gamma)}
\def\keng{K^+ \to e^+ \nu_{e} (\gamma)}
\def\kppp{K^+ \to \pi^+ \pi^{+,0} \pi^{-,0}}
\def\kpeng{K^+ \to \pi^0 e^+ \nu_e (\gamma)}
\def\kpmng{K^+ \to \pi^0 \mu^+ \nu_{\mu} (\gamma)}
\def\kppeng{K^+ \to \pi^{+,0} \pi^{-,0} e^+ \nu_e (\gamma)}
\def\kppmng{K^+ \to \pi^{+,0} \pi^{-,0} \mu^+ \nu_{\mu} (\gamma)}

\usepackage[backend=bibtex,style=numeric-comp,natbib=true,sorting=none,block=none,citestyle=nature]{biblatex} 
\usepackage[babel]{csquotes} 
\begin{document}
\pagenumbering{arabic}
\centerline{\LARGE EUROPEAN ORGANIZATION FOR NUCLEAR RESEARCH}
\vspace{15mm}
\begin{flushright}
CERN-EP-2020-193\\
\today \\
 \vspace{2mm}
\end{flushright}
\vspace{15mm}

\begin{center}
\Large{\bf Search for \boldmath{$\pi^0$} decays to invisible particles \\
\vspace{5mm}
}
The NA62 Collaboration
\end{center}
\vspace{10mm}
\begin{abstract}
The NA62 experiment at the CERN SPS reports a study of a sample of $4 \times10^{9}$ tagged $\pi^0$ mesons from $\kppg$, searching for the decay of the $\pi^0$ to invisible particles. No signal is observed in excess of the expected background fluctuations. An upper limit of $4.4 \times10^{-9}$ is set on the branching ratio at 90\% confidence level, improving on previous results by a factor of 60. 
This result can also be interpreted as a model-independent upper limit on the branching ratio for the decay $K^+ \to \pi^+ X$, where $X$ is a particle escaping detection with mass in the range 0.110--0.155~GeV$/c^2$ and rest lifetime greater than 100~ps. 
Model-dependent upper limits are obtained assuming $X$ to be an axion-like particle with dominant fermion couplings or a dark scalar mixing with the Standard Model Higgs boson.
\end{abstract}
\vspace{20mm}

\begin{center}
\em{To be submitted to JHEP}
\end{center}

\clearpage
\begin{center}
{\Large The NA62 Collaboration$\,$\renewcommand{\thefootnote}{\fnsymbol{footnote}}%
\footnotemark[1]\renewcommand{\thefootnote}{\arabic{footnote}}}\\
\end{center}
 \vspace{3mm}
\begin{raggedright}
\noindent

{\bf Universit\'e Catholique de Louvain, Louvain-La-Neuve, Belgium}\\
 E.~Cortina Gil,
 A.~Kleimenova,
 E.~Minucci$\,$\footnotemark[1]$^,\,$\footnotemark[2],
 S.~Padolski$\,$\footnotemark[3],
 P.~Petrov,
 A.~Shaikhiev$\,$\footnotemark[4],
 R.~Volpe$\,$\footnotemark[5]\\[2mm]

{\bf TRIUMF, Vancouver, British Columbia, Canada}\\
 T.~Numao,
 B.~Velghe\\[2mm]

{\bf University of British Columbia, Vancouver, British Columbia, Canada}\\
 D.~Bryman$\,$\footnotemark[6],
 J.~Fu$\,$\footnotemark[7]\\[2mm]

{\bf Charles University, Prague, Czech Republic}\\
 T.~Husek$\,$\footnotemark[8],
 J.~Jerhot$\,$\footnotemark[9],
 K.~Kampf,
 M.~Zamkovsky\\[2mm]

{\bf Institut f\"ur Physik and PRISMA Cluster of excellence, Universit\"at Mainz, Mainz, Germany}\\
 R.~Aliberti$\,$\footnotemark[10],
 G.~Khoriauli$\,$\footnotemark[11],
 J.~Kunze,
 D.~Lomidze$\,$\footnotemark[12],
 R.~Marchevski$\,$\footnotemark[13],
 L.~Peruzzo$\renewcommand{\thefootnote}{\fnsymbol{footnote}}\footnotemark[1]\renewcommand{\thefootnote}{\arabic{footnote}}$,
 M.~Vormstein,
 R.~Wanke\\[2mm]

{\bf Dipartimento di Fisica e Scienze della Terra dell'Universit\`a e INFN, Sezione di Ferrara, Ferrara, Italy}\\
 P.~Dalpiaz,
 M.~Fiorini,
 I.~Neri,
 A.~Norton,
 F.~Petrucci,
 H.~Wahl\\[2mm]

{\bf INFN, Sezione di Ferrara, Ferrara, Italy}\\
 A.~Cotta Ramusino,
 A.~Gianoli\\[2mm]

{\bf Dipartimento di Fisica e Astronomia dell'Universit\`a e INFN, Sezione di Firenze, Sesto Fiorentino, Italy}\\
 E.~Iacopini,
 G.~Latino,
 M.~Lenti,
 A.~Parenti\\[2mm]

{\bf INFN, Sezione di Firenze, Sesto Fiorentino, Italy}\\
 A.~Bizzeti$\,$\footnotemark[14],
 F.~Bucci\\[2mm]

{\bf Laboratori Nazionali di Frascati, Frascati, Italy}\\
 A.~Antonelli,
 G.~Georgiev$\,$\footnotemark[15],
 V.~Kozhuharov$\,$\footnotemark[15],
 G.~Lanfranchi,
 S.~Martellotti,
 M.~Moulson,
 T.~Spadaro$\renewcommand{\thefootnote}{\fnsymbol{footnote}}\footnotemark[1]\renewcommand{\thefootnote}{\arabic{footnote}}$\\[2mm]

{\bf Dipartimento di Fisica ``Ettore Pancini'' e INFN, Sezione di Napoli, Napoli, Italy}\\
 F.~Ambrosino,
 T.~Capussela,
 M.~Corvino$\,$\footnotemark[13],
 D.~Di Filippo,
 P.~Massarotti,
 M.~Mirra,
 M.~Napolitano,
 G.~Saracino\\[2mm]

{\bf Dipartimento di Fisica e Geologia dell'Universit\`a e INFN, Sezione di Perugia, Perugia, Italy}\\
 G.~Anzivino,
 F.~Brizioli,
 E.~Imbergamo,
 R.~Lollini,
 R.~Piandani$\,$\footnotemark[16],
 C.~Santoni\\[2mm]

{\bf INFN, Sezione di Perugia, Perugia, Italy}\\
 M.~Barbanera$\,$\footnotemark[17],
 P.~Cenci,
 B.~Checcucci,
 P.~Lubrano,
 M.~Lupi$\,$\footnotemark[18],
 M.~Pepe,
 M.~Piccini\\[2mm]

{\bf Dipartimento di Fisica dell'Universit\`a e INFN, Sezione di Pisa, Pisa, Italy}\\
 F.~Costantini,
 L.~Di Lella,
 N.~Doble,
 M.~Giorgi,
 S.~Giudici,
 G.~Lamanna,
 E.~Lari,
 E.~Pedreschi,
 M.~Sozzi\\[2mm]

{\bf INFN, Sezione di Pisa, Pisa, Italy}\\
 C.~Cerri,
 R.~Fantechi,
 L.~Pontisso,
 F.~Spinella\\[2mm]
\newpage
{\bf Scuola Normale Superiore e INFN, Sezione di Pisa, Pisa, Italy}\\
 I.~Mannelli\\[2mm]

{\bf Dipartimento di Fisica, Sapienza Universit\`a di Roma e INFN, Sezione di Roma I, Roma, Italy}\\
 G.~D'Agostini,
 M.~Raggi\\[2mm]

{\bf INFN, Sezione di Roma I, Roma, Italy}\\
 A.~Biagioni,
 E.~Leonardi,
 A.~Lonardo,
 P.~Valente,
 P.~Vicini\\[2mm]

{\bf INFN, Sezione di Roma Tor Vergata, Roma, Italy}\\
 R.~Ammendola,
 V.~Bonaiuto$\,$\footnotemark[19],
 A.~Fucci,
 A.~Salamon,
 F.~Sargeni$\,$\footnotemark[20]\\[2mm]

{\bf Dipartimento di Fisica dell'Universit\`a e INFN, Sezione di Torino, Torino, Italy}\\
 R.~Arcidiacono$\,$\footnotemark[21],
 B.~Bloch-Devaux,
 M.~Boretto$\,$\footnotemark[13],
 E.~Menichetti,
 E.~Migliore,
 D.~Soldi\\[2mm]

{\bf INFN, Sezione di Torino, Torino, Italy}\\
 C.~Biino,
 A.~Filippi,
 F.~Marchetto\\[2mm]

{\bf Instituto de F\'isica, Universidad Aut\'onoma de San Luis Potos\'i, San Luis Potos\'i, Mexico}\\
 J.~Engelfried,
 N.~Estrada-Tristan$\,$\footnotemark[22]\\[2mm]

{\bf Horia Hulubei National Institute of Physics for R\&D in Physics and Nuclear Engineering, Bucharest-Magurele, Romania}\\
 A. M.~Bragadireanu,
 S. A.~Ghinescu,
 O. E.~Hutanu\\[2mm]

{\bf Joint Institute for Nuclear Research, Dubna, Russia}\\
 A.~Baeva,
 D.~Baigarashev,
 D.~Emelyanov,
 T.~Enik,
 V.~Falaleev,
 V.~Kekelidze,
 A.~Korotkova,
 L.~Litov$\,$\footnotemark[15],
 D.~Madigozhin,
 M.~Misheva$\,$\footnotemark[23],
 N.~Molokanova,
 S.~Movchan,
 I.~Polenkevich,
 Yu.~Potrebenikov,
 S.~Shkarovskiy,
 A.~Zinchenko$\,$\renewcommand{\thefootnote}{\fnsymbol{footnote}}\footnotemark[2]\renewcommand{\thefootnote}{\arabic{footnote}}\\[2mm]

{\bf Institute for Nuclear Research of the Russian Academy of Sciences, Moscow, Russia}\\
 S.~Fedotov,
 E.~Gushchin,
 A.~Khotyantsev,
 Y.~Kudenko$\,$\footnotemark[24],
 V.~Kurochka,
 M.~Medvedeva,
 A.~Mefodev\\[2mm]

{\bf Institute for High Energy Physics - State Research Center of Russian Federation, Protvino, Russia}\\
 S.~Kholodenko,
 V.~Kurshetsov,
 V.~Obraztsov,
 A.~Ostankov$\,$\renewcommand{\thefootnote}{\fnsymbol{footnote}}\footnotemark[2]\renewcommand{\thefootnote}{\arabic{footnote}},
 V.~Semenov$\,$\renewcommand{\thefootnote}{\fnsymbol{footnote}}\footnotemark[2]\renewcommand{\thefootnote}{\arabic{footnote}},
 V.~Sugonyaev,
 O.~Yushchenko\\[2mm]

{\bf Faculty of Mathematics, Physics and Informatics, Comenius University, Bratislava, Slovakia}\\
 L.~Bician$\,$\footnotemark[13],
 T.~Blazek,
 V.~Cerny,
 Z.~Kucerova\\[2mm]

{\bf CERN,  European Organization for Nuclear Research, Geneva, Switzerland}\\
 J.~Bernhard,
 A.~Ceccucci,
 H.~Danielsson,
 N.~De Simone$\,$\footnotemark[25],
 F.~Duval,
 B.~D\"obrich,
 L.~Federici,
 E.~Gamberini,
 L.~Gatignon,
 R.~Guida,
 F.~Hahn$\,$\renewcommand{\thefootnote}{\fnsymbol{footnote}}\footnotemark[2]\renewcommand{\thefootnote}{\arabic{footnote}},
 E. B.~Holzer,
 B.~Jenninger,
 M.~Koval$\,$\footnotemark[26],
 P.~Laycock$\,$\footnotemark[3],
 G.~Lehmann Miotto,
 P.~Lichard,
 A.~Mapelli,
 K.~Massri,
 M.~Noy,
 V.~Palladino$\,$\footnotemark[27],
 M.~Perrin-Terrin$\,$\footnotemark[28]$^,\,$\footnotemark[29],
 J.~Pinzino$\,$\footnotemark[30]$^,\,$\footnotemark[31],
 V.~Ryjov,
 S.~Schuchmann$\,$\footnotemark[32],
 S.~Venditti\\[2mm]

{\bf University of Birmingham, Birmingham, United Kingdom}\\
 T.~Bache,
 M. B.~Brunetti$\,$\footnotemark[33],
 V.~Duk$\,$\footnotemark[34],
 V.~Fascianelli$\,$\footnotemark[35],
 J. R.~Fry,
 F.~Gonnella,
 E.~Goudzovski,
 L.~Iacobuzio,
 C.~Lazzeroni,
 N.~Lurkin$\,$\footnotemark[9],
 F.~Newson,
 C.~Parkinson$\,$\footnotemark[9],
 A.~Romano,
 A.~Sergi,
 A.~Sturgess,
 J.~Swallow\\[2mm]
\newpage
{\bf University of Bristol, Bristol, United Kingdom}\\
 H.~Heath,
 R.~Page,
 S.~Trilov\\[2mm]

{\bf University of Glasgow, Glasgow, United Kingdom}\\
 B.~Angelucci,
 D.~Britton,
 C.~Graham,
 D.~Protopopescu\\[2mm]

{\bf University of Lancaster, Lancaster, United Kingdom}\\
 J.~Carmignani,
 J. B.~Dainton,
 R. W. L.~Jones,
 G.~Ruggiero\\[2mm]

{\bf University of Liverpool, Liverpool, United Kingdom}\\
 L.~Fulton,
 D.~Hutchcroft,
 E.~Maurice$\,$\footnotemark[36],
 B.~Wrona\\[2mm]

{\bf George Mason University, Fairfax, Virginia, USA}\\
 A.~Conovaloff,
 P.~Cooper,
 D.~Coward$\,$\footnotemark[37],
 P.~Rubin 

\end{raggedright}
%
%
\setcounter{footnote}{0}
\renewcommand{\thefootnote}{\fnsymbol{footnote}}
\footnotetext[1]{Corresponding authors: L.~Peruzzo, T.~Spadaro, 
email: letizia.peruzzo@cern.ch, tommaso.spadaro@cern.ch}
\footnotetext[2]{Deceased}
\renewcommand{\thefootnote}{\arabic{footnote}}

\footnotetext[1]{Present address: Laboratori Nazionali di Frascati, I-00044 Frascati, Italy}
\footnotetext[2]{Also at CERN,  European Organization for Nuclear Research, CH-1211 Geneva 23, Switzerland}
\footnotetext[3]{Present address: Brookhaven National Laboratory, Upton, NY 11973, USA}
\footnotetext[4]{Also at Institute for Nuclear Research of the Russian Academy of Sciences, 117312 Moscow, Russia}
\footnotetext[5]{Present address: Faculty of Mathematics, Physics and Informatics, Comenius University, 842 48, Bratislava, Slovakia}
\footnotetext[6]{Also at TRIUMF, Vancouver, British Columbia, V6T 2A3, Canada}
\footnotetext[7]{Present address: UCLA Physics and Biology in Medicine, Los Angeles, CA 90095, USA}
\footnotetext[8]{Present address: Department of Astronomy and Theoretical Physics, Lund University, Lund, SE 223-62, Sweden}
\footnotetext[9]{Present address: Universit\'e Catholique de Louvain, B-1348 Louvain-La-Neuve, Belgium}
\footnotetext[10]{Present address: Institut f\"ur Kernphysik and Helmholtz Institute Mainz, Universit\"at Mainz, Mainz, D-55099, Germany}
\footnotetext[11]{Present address: Universit\"at W\"urzburg, D-97070 W\"urzburg, Germany}
\footnotetext[12]{Present address: Universit\"at Hamburg, D-20146 Hamburg, Germany}
\footnotetext[13]{Present address: CERN,  European Organization for Nuclear Research, CH-1211 Geneva 23, Switzerland}
\footnotetext[14]{Also at Dipartimento di Fisica, Universit\`a di Modena e Reggio Emilia, I-41125 Modena, Italy}
\footnotetext[15]{Also at Faculty of Physics, University of Sofia, BG-1164 Sofia, Bulgaria}
\footnotetext[16]{Present address: Institut f\"ur Experimentelle Teilchenphysik (KIT), D-76131 Karlsruhe, Germany}
\footnotetext[17]{Present address: School of Physics and Astronomy, University of Birmingham, Birmingham, B15 2TT, UK}
\footnotetext[18]{Present address: Institut am Fachbereich Informatik und Mathematik, Goethe Universit\"at, D-60323 Frankfurt am Main, Germany}
\footnotetext[19]{Also at Department of Industrial Engineering, University of Roma Tor Vergata, I-00173 Roma, Italy}
\footnotetext[20]{Also at Department of Electronic Engineering, University of Roma Tor Vergata, I-00173 Roma, Italy}
\footnotetext[21]{Also at Universit\`a degli Studi del Piemonte Orientale, I-13100 Vercelli, Italy}
\footnotetext[22]{Also at Universidad de Guanajuato, Guanajuato, Mexico}
\footnotetext[23]{Present address: Institute of Nuclear Research and Nuclear Energy of Bulgarian Academy of Science (INRNE-BAS), BG-1784 Sofia, Bulgaria}
\footnotetext[24]{Also at National Research Nuclear University (MEPhI), 115409 Moscow and Moscow Institute of Physics and Technology, 141701 Moscow region, Moscow, Russia}
\footnotetext[25]{Present address: DESY, D-15738 Zeuthen, Germany}
\footnotetext[26]{Present address: Charles University, 116 36 Prague 1, Czech Republic}
\footnotetext[27]{Present address: Physics Department, Imperial College London, London, SW7 2BW, UK}
\footnotetext[28]{Present address: Aix Marseille University, CNRS/IN2P3, CPPM, F-13288, Marseille, France}
\footnotetext[29]{Also at Universit\'e Catholique de Louvain, B-1348 Louvain-La-Neuve, Belgium}
\footnotetext[30]{Present address: Department of Physics, University of Toronto, Toronto, Ontario, M5S 1A7, Canada}
\footnotetext[31]{Also at INFN, Sezione di Pisa, I-56100 Pisa, Italy}
\footnotetext[32]{Present address: Institut f\"ur Physik and PRISMA Cluster of excellence, Universit\"at Mainz, D-55099 Mainz, Germany}
\footnotetext[33]{Present address: Department of Physics, University of Warwick, Coventry, CV4 7AL, UK}
\footnotetext[34]{Present address: INFN, Sezione di Perugia, I-06100 Perugia, Italy}
\footnotetext[35]{Present address: Dipartimento di Psicologia, Universit\`a di Roma La Sapienza, I-00185 Roma, Italy}
\footnotetext[36]{Present address: Laboratoire Leprince Ringuet, F-91120 Palaiseau, France}
\footnotetext[37]{Also at SLAC National Accelerator Laboratory, Stanford University, Menlo Park, CA 94025, USA}

\clearpage
\section{Introduction}
\label{sec:intro}
The relevance of the decay $\pi^0 \to \nu \bar{\nu}$ was first highlighted in the study of neutrino properties such as mass,
helicity and number of families~\cite{RefPi0NuNuPaper1,RefPi0NuNuPaper2}.
The decay is forbidden for pure left-handed massless neutrinos due to angular momentum conservation.
The observation of neutrino oscillations demonstrated the existence of non-zero neutrino masses, allowing the $\pi^0 \to \nu \bar{\nu}$ decay via $Z$-boson exchange.
The direct experimental limit on the tau neutrino mass, $m_{\nu_{\tau}} < 18.2$~MeV/$c^2$ at 95\% confidence level (CL)~\cite{RefNuTauMassPaper}, corresponds to a branching ratio ($\mathrm{BR}$) for $\pi^0 \to \nu \bar{\nu}$ below $5 \times 10^{-10}$ at 90\% CL. 
A more stringent limit is set by cosmological constraints on the sum of the neutrino masses: $\Sigma m_{\nu} \lesssim 1$~eV$/c^2$, implying $\mathrm{BR} (\pi^0 \to \nu \bar{\nu})  < 10^{-24}$, which is well below the current experimental sensitivity.
The BR for the decay $\pi^0\to\nu\overline{\nu}\nu\overline{\nu}$ is expected to be well below the current experimental sensitivity, too~\cite{DaoNeng}.
The current 
experimental limit is $2.7 \times 10^{-7}$ at 90\% CL~\cite{RefBNLE949Paper}. 

Beyond determining neutrino properties, the search for a $\pi^0$ decay to any invisible final state (``$\dec$'' in the following) allows tests of new-physics scenarios involving feebly-interacting long-lived particles. 
Any observation of $\dec$ at the currently available sensitivity would be an indication of new physics.
The results of the analysis presented here may also be interpreted as a search for the production of axion-like particles (ALPs) with dominant fermion couplings or dark scalar particles from $K^+$ decays~\cite{RefAxion_DMTaste,RefPBC_Document}.

The search for $\dec$ is performed with tagged $\pi^0$ mesons from the decay chain
\begin{equation}
\label{eq:decaychain}
\kppg,\quad \dec \text{,}
\end{equation}
inclusive of final states with additional photons from radiative processes. 

An abundant flux of $K^+$ mesons is provided by a 75~GeV/$c$ unseparated hadron beam from the CERN Super Proton Synchrotron (SPS). The search is performed with NA62, a fixed target experiment operating in the SPS North Area with the main goal of measuring the BR of the ultra-rare decay $K^+ \rightarrow \pi^+ \nu \bar\nu$. 
The design of the experiment enables operation at high beam intensity, hermetic photon veto coverage, full particle identification and high-rate tracking with low material budget.
The NA62 detector has been fully operational since 2016.
The results reported here are obtained from the analysis of a data sample collected in 2017.
By exploiting the performance of the NA62 photon-veto system, a rejection of $\mathcal{O}(10^{8}\text{--}10^{9})$ for any visible $\pi^0$ decay in $\kppg$ events is obtained for $\pi^+$ momenta in the range 10--45~GeV/$c$.
%
\section{Beam line and detector}
\label{sec:layout}
\begin{figure}[!ht]
\center
\includegraphics[width=1.08\textwidth]{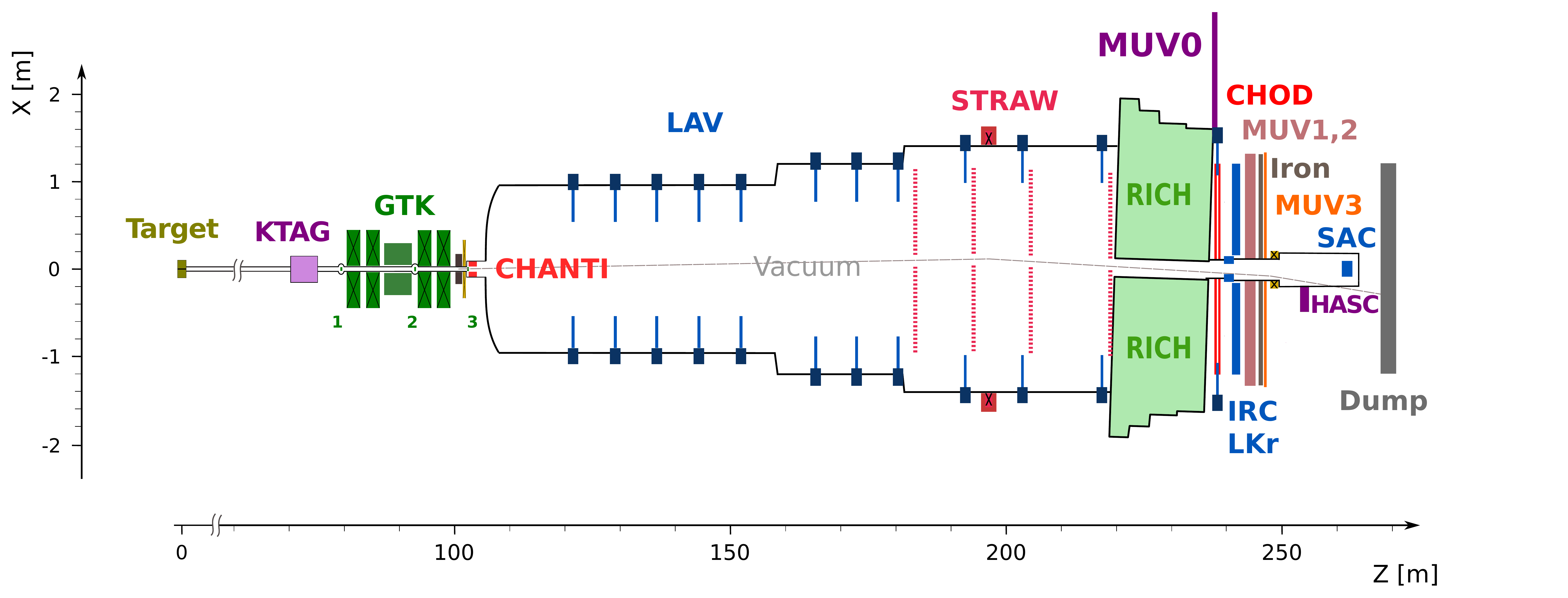}
\caption{Schematic top view of the NA62 beam line and detector. Dipole magnets are displayed as boxes with superimposed crosses. The ``CHOD'' label indicates both the CHOD and NA48-CHOD hodoscopes described in the text. Also shown is the trajectory of a beam particle in vacuum which crosses all the detector apertures, thus avoiding interactions with material. A dipole magnet between MUV3 and SAC deflects the beam particles out of the SAC acceptance.}
\label{fig:NA62layout}
\end{figure}

The NA62 beam line and detector are shown in Figure~\ref{fig:NA62layout}, with a detailed description presented in~\cite{RefNA62DetectorPaper}.
The beam line defines the Z-axis of the experiment's right-handed coordinate system. The origin is the kaon production target, and beam particles travel in the positive Z-direction. The Y-axis is vertical (positive up), and the X-axis is horizontal. 

The kaon production target is a 40~cm long beryllium rod. A 400~GeV proton beam extracted from the CERN SPS impinges on the target in spills of 
3~s
effective duration. Typical intensities during data taking range from  $1.7$ to $1.9\times10^{12}$ protons per pulse. The resulting secondary hadron beam of positively charged particles consists of 70\% $\pi^+$, 23\% protons, and 6\% $K^+$, with a nominal momentum of 75~GeV/$c$ and 1\% rms momentum spread. 

Beam particles are characterized by a differential Cherenkov counter (KTAG) and a three-station silicon pixel array (Gigatracker, GTK). The KTAG uses N$_2$ gas at 1.75~bar pressure (contained in a 5~m long vessel) and is read out by photomultiplier tubes grouped in eight sectors. It tags incoming kaons with 70~ps time resolution. The GTK stations are located before, between, and after two pairs of dipole magnets (a beam achromat), forming a spectrometer that measures beam particle momentum, direction, and time with resolutions of 0.15~GeV/$c$, 16~$\mu$rad, and 100~ps, respectively.
The typical beam particle rate at the third GTK station (GTK3) is 450~MHz.
This last station is immediately preceded by a 1~m thick, variable-aperture, steel collimator.  Its inner aperture is typically set at
66~mm~$\times$~33~mm, and its outer dimensions are about 150~mm.  It serves as a partial shield against hadrons produced by upstream $K^+$ decays.

GTK3 marks the beginning of a 117~m long vacuum tank. The first 80~m of the tank define a volume in which 13\% of the kaons decay. The beam has a  rectangular transverse profile of $52 \times 24$~mm$^2$ and a divergence of 0.11~mrad (rms) in each plane at the decay volume entrance.

The time, momentum, and direction of charged particles emitted by in-flight kaon decays are measured by a magnetic spectrometer (STRAW), a ring-imaging Cherenkov counter (RICH), and two adjacent scintillator hodoscopes (CHOD and NA48-CHOD). The STRAW, consisting of two pairs of straw chambers on either side of a dipole magnet, measures momentum vectors with a resolution $\sigma_p / p$ between 0.3\% and 0.4\%.
The angular resolution decreases from 60~$\mu$rad at 10~GeV/$c$ to 20~$\mu$rad at 50~GeV/$c$ momentum.
The RICH, filled with neon at atmospheric pressure, tags the decay particles with a timing precision of better than 100~ps and provides particle identification. The CHOD, a matrix of tiles read out by SiPMs, and the NA48-CHOD, composed of two orthogonal planes of scintillating slabs reused from the NA48 experiment, are used for triggering and timing, providing a time measurement with 200~ps resolution.
 
Other sub-detectors suppress decays into photons or multiple charged particles (electrons, pions or muons) or provide complementary particle identification. Six stations of plastic scintillator bars (CHANTI) detect extra activity, including inelastic interactions in GTK3, with 99\% efficiency and 1~ns time resolution. Twelve stations of ring-shaped veto detectors
(LAV1 to LAV12), made of lead-glass blocks, surround the vacuum tank and downstream sub-detectors to achieve hermetic acceptance for photons emitted by $K^+$ decays in the decay volume at polar angles between 10 and 50~mrad. A 27 radiation-length, quasi-homogeneous liquid krypton electromagnetic calorimeter (LKr) detects photons from $K^+$ decays emitted at angles between 1 and 10~mrad. The LKr provides particle identification information complementary to that of the RICH. The LKr energy resolution is $\sigma_E / E = 1.4\%$ for energy deposits of 25~GeV. Its spatial and time resolutions are 1~mm and between 0.5 and 1~ns, respectively, depending on the amount and type of energy released. Two hadronic iron/scintillator-strip sampling calorimeters (MUV1,2) and an array of scintillator tiles located behind 80~cm of iron (MUV3) supplement the pion/muon identification system. MUV3 has a time resolution of 400~ps. A lead/scintillator shashlik calorimeter (IRC) located in front of the LKr, covering an annular region between 65 and 135~mm from the Z-axis, and a similar detector (SAC) placed on the Z-axis at the downstream end of the apparatus, ensure the detection of photons down to zero degrees in the forward direction. Additional counters (MUV0, HASC) are installed at optimized locations to extend coverage for charged particles produced in multi-track kaon decays. 

All detectors are read out with time-to-digital converters (TDCs), except for LKr and MUV1,2, which are read out with 14-bit flash analog-to-digital converters. The IRC and SAC are read out with both. All TDCs are mounted on custom-made (TEL62) boards, except for GTK and STRAW, which each have specialized TDC boards. TEL62 boards both read out data and provide trigger information. A dedicated processor interprets LKr calorimeter signals for triggering.
A custom-made board combines logic signals (primitives) from the RICH, CHOD, NA48-CHOD, LKr, LAV, and MUV3 into a low-level trigger (L0) whose decision is dispatched to sub-detectors for data readout~\cite{RefNA62TDAQ}. A software trigger (L1) exploits reconstruction algorithms similar to those used offline with data from KTAG, LAV, and STRAW to further reduce the data before storing it on disk. 

The L0 trigger condition used to search for the decay chain of Equation~(\ref{eq:decaychain}), referred to as the $\pi\nu\bar{\nu}$ trigger, aims to select final states with one emitted $\pi^+$ and missing energy. It requires a primitive in the RICH in coincidence within 10~ns with a primitive from at least one CHOD tile. 
No primitive deriving from signals in opposite CHOD quadrants must be found within the 10~ns window, reducing the contribution of $K^+\to\pi^+\pi^+\pi^-$ decays and in general of final states with multiple charged particles. No in-time primitive from MUV3 must be present, reducing the contribution of $K^+\to(\pi^{0})\mu^{+}\nu$ decays. 
In the LKr no more than one energy deposit above 1~GeV must be found within 6.5~ns of the RICH primitive and the total reconstructed energy is required to be below 30~GeV.
These conditions reduce the contribution of multi-photon final states and are particularly effective in rejecting forward-boosted photons from $\pi^0$ decays.
Additional conditions are applied at L1 to further ensure the presence of a charged kaon decaying to a single charged particle, while rejecting final states with additional particles emitted at large angle. The charged kaon must be positively identified using KTAG information within 10~ns of the L0 trigger time derived from the RICH. At least one track must be reconstructed with the STRAW as a particle with momentum below 50~GeV/$c$ and forming a vertex with the nominal beam axis upstream of the first STRAW chamber. Events with in-time signals from three or more LAV blocks are rejected.  
These conditions reduce the trigger rate by a factor of 100.

The analysis also uses data collected with a minimum-bias L0 trigger (control trigger) based on NA48-CHOD information, downscaled by a factor of 400, for signal normalization, efficiency measurement, and background estimation.  

The detector response is simulated with a Monte Carlo simulation programme based on the GEANT4 package~\cite{GEANT4}. The main $K^+$ decay modes are simulated including final states with additional photons from radiative processes.
\section{Analysis principle}
\label{sec:principle}
A high-statistics sample of tagged $\pi^0$ mesons is obtained by the reconstruction of the charged particles in the process $\kppg$.
The $K^+$ track, called the beam track, is reconstructed with the GTK spectrometer; the $\pi^+$ track, called the downstream track, is reconstructed with the STRAW spectrometer. 
The beam and downstream tracks must be positively identified as a $K^+$ and a $\pi^+$, have consistent timing and form a vertex inside a defined fiducial volume.
Kaon and pion energies are obtained from the measured kaon and pion momenta assuming the PDG mass values~\cite{RefPDG}.
The two-body kinematics of the $\pi^+ \pi^0$ final state is enforced by the requirement that the squared missing mass computed from the kaon and pion 4-momenta ($P_{K^+}$ and $P_{\pi^+}$),
\begin{equation}
M^{2}_{\mathrm{miss}} = \left( P_{K^+} - P_{\pi^+} \right) ^2,
\label{eq:mmiss}
\end{equation}
be compatible with the squared $\pi^0$ mass~\cite{RefPDG}.
Requirements on the multiplicity of charged tracks efficiently reject events with $\pi^0$ decays other than $\pi^0\to \gamma\gamma$. This selection is applied to control-trigger data to determine, after the correction for the trigger downscaling factor, the number of tagged $\pi^0$ mesons $N_{\pi^0}$ used for normalization. 

Further requirements are applied to $\pi\nu\bar{\nu}$-trigger data to reject events with any additional activity correlated in time with the incoming $K^+$ and its daughter $\pi^+$. The branching ratio for $\dec$ is computed as:
\begin{equation}
\mathrm{BR} (\dec) = \mathrm{BR} (\pi^0 \to \gamma \gamma) \times \frac{N_\mathrm{s}}{N_{\pi^0} \times \varepsilon_{\mathrm{sel}} \times \varepsilon_{\mathrm{trig}}},
\label{eq:BrPi0Invisible}
\end{equation}
where $N_\mathrm{s}$ is the number of signal events, obtained after background subtraction from the observed number of candidate events, and $\varepsilon_{\mathrm{sel}}$ and $\varepsilon_{\mathrm{trig}}$ are the efficiencies of the signal-selection algorithm and the $\pi\nu\bar{\nu}$ trigger, respectively.
In Equation~(\ref{eq:BrPi0Invisible}), the efficiency for the selection of the normalization sample cancels when considering the ratio of $\pi^0\to\gamma\gamma$ and $\pi^0\to\mathrm{invisible}$, while the control-trigger efficiency for $\pi^0\to\gamma\gamma$ is equal to one to within a few parts per thousand.

After the signal-sample selection, the residual background is dominated by $\kppg$ events with $\pi^0 \to \gamma \gamma$ where all the photons are undetected. The rejection power against this background is estimated \textit{a priori} from a combination of data-based studies and Monte Carlo (MC) simulations. The single-photon efficiencies of the sub-detectors composing the photon-veto system are evaluated in control-trigger data and MC simulation samples.
The detector efficiencies are combined to evaluate the expected background rejection, which strongly depends on the $\pi^+$ momentum, with a sample of simulated $\kppg$ events. 
The analysis is performed in a range of the $\pi^+$ momentum chosen to optimize the sensitivity in the no-signal hypothesis. The data sample is kept masked until validation of the background evaluation. The validation is performed with data sub-samples in which no signal is expected on the basis of results from previous searches. 
\section{Selection of the normalization sample}
\label{sec:normSelection}
The normalization sample consists of $\kppg$ events selected from the control-trigger data sample by the use of information from only the $K^+$ and $\pi^+$ particles.
The selection criteria require the presence of a good-quality, positively charged downstream track, in-time and geometrically associated with activity in the NA48-CHOD, CHOD, LKr, RICH and MUV1,2 detectors. The time of the downstream track is evaluated with a resolution at the level of 100~ps by combination of the STRAW, NA48-CHOD, LKr, and RICH time measurements.

Information from the LKr and MUV1,2 calorimeters and from the MUV3 and RICH detectors are used to achieve high-purity pion identification: no in-time MUV3 activity must be found geometrically associated with the downstream track; the spatial energy profiles in the LKr and MUV1,2 calorimeters are combined into a multivariate classifier trained to distinguish charged pions from muons or positrons and a minimum probability for $\pi^+$ is required; finally, the Cherenkov ring associated with the downstream track must be compatible with that expected from a charged pion. The identification of $\pi^+$ mesons is achieved with an efficiency around 70\% and a muon contamination of the order of $10^{-6}$ for momenta in the range 10--50~GeV$/c$.

The identification of a $K^+$ meson from the beam is ensured by the requirement of signals in time with the downstream track from at least five KTAG sectors. The momentum and direction of the $K^+$ are measured with the GTK.
For any GTK track, the time difference with respect to the KTAG signal and the closest distance of approach (CDA) to the downstream track are combined into a likelihood variable. The GTK track with the highest likelihood is selected, provided it satisfies stringent quality criteria, and called beam track. The beam track momentum must lie within the range 72--78~GeV/$c$ and its direction is required to be consistent with the distribution expected for beam particles. 

If more than one downstream track reaches this stage of the selection, the one closest in time to the trigger is selected.
The vertex is defined as the point of closest approach between the downstream and beam tracks.
The vertex is required to lie in a fiducial volume (FV) beginning 105~m downstream of the target and extending for 60~m along the Z-axis.

Beam particles decaying upstream of the decay volume or interacting in the GTK material can produce secondaries leading to an incorrect $K / \pi$ association. Additional requirements are applied to reject such beam-related backgrounds. The most relevant conditions are: no in-time activity must be present in the CHANTI detector; the backward extrapolation of the downstream track to the plane Z~$=102$~m must be outside a beam-activity region defined as $|\mathrm{X}| \leq 100$~mm and $|\mathrm{Y}| \leq 500$~mm; 
the longitudinal position of the vertex and the position of the downstream track at the first STRAW chamber are required to be in a region determined by the distribution of the $\pi^+$ angle of emission for real $K^+$ decays.
These criteria reduce the contamination from beam-related background to less than $10^{-5}$, as determined from data control samples~\cite{RefNA62Pnn2017}.

\begin{figure}[htb]
\centering
\includegraphics[width=0.7\textwidth]{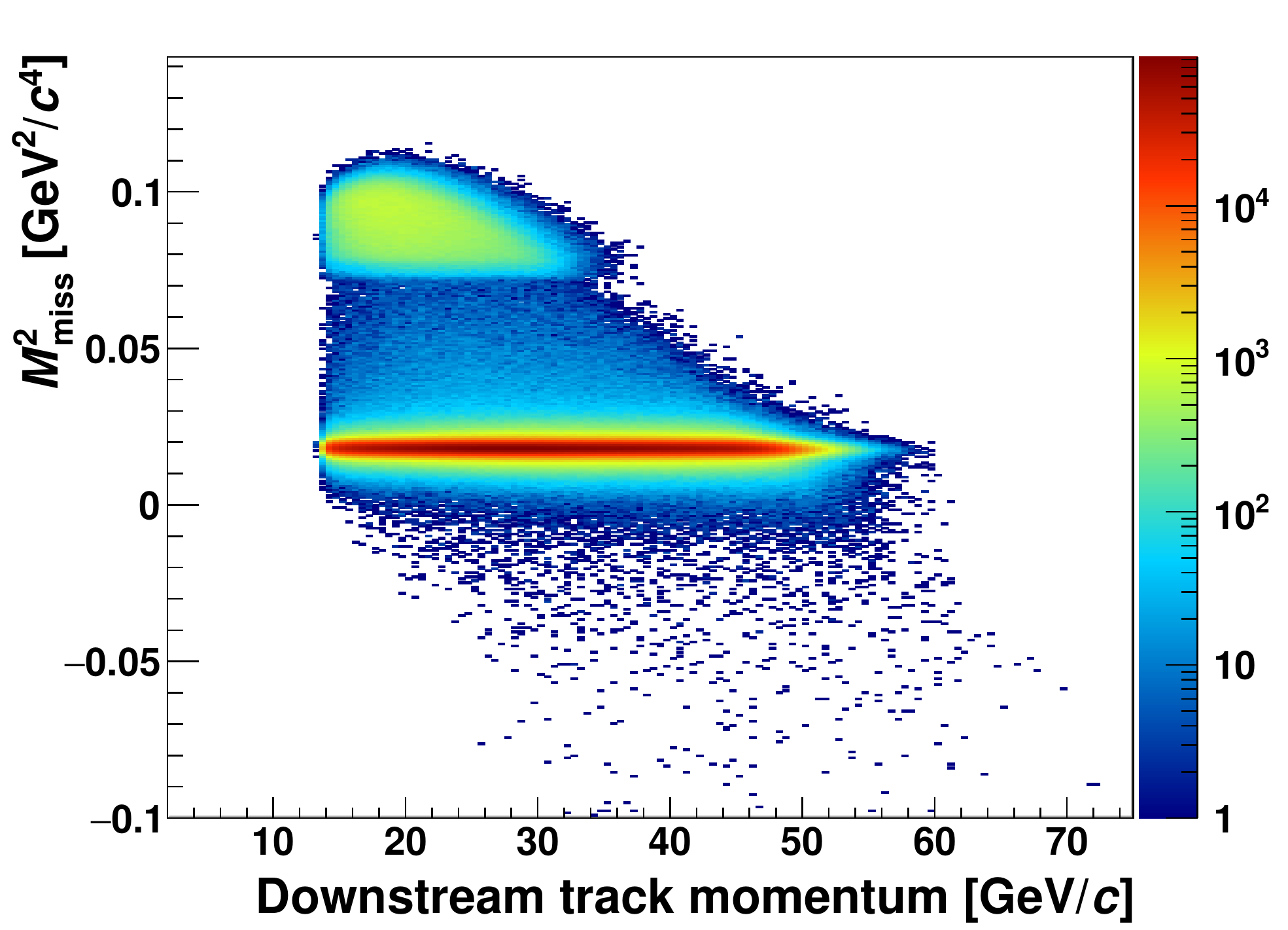}
\caption{Distribution of the squared missing mass versus the downstream track momentum for control-trigger data, where the selection algorithm for the normalization sample is applied except for the conditions of Equations~(\ref{eq:mmissCut}) and~(\ref{eq:mmissCut2}). 
}
\label{fig:normalizationSample}
\end{figure}

Using the measured beam and downstream track momenta, the squared missing mass $M^{2}_{\mathrm{miss}}$ is evaluated with mass assignments as in Equation~(\ref{eq:mmiss}). The distribution of $M^{2}_{\mathrm{miss}}$ as a function of the downstream track momentum, shown in Figure~\ref{fig:normalizationSample}, is dominated by $\kppg$ decays peaking around the squared $\pi^0$ mass with a resolution of $10^{-3}$~GeV$^{2}/c^{4}$. Events with $K^+\to\pi^+\pi^0\pi^0$ decays populate the region $M^{2}_{\mathrm{miss}} > 0.07$~$\mathrm{GeV}^2/c^4$. The following condition is required:
\begin{equation}
\label{eq:mmissCut}
0.015 < M^{2}_{\mathrm{miss}} <  0.021\text{~GeV}^{2}/c^4.
\end{equation}
To refine the kinematic identification of the $\kppg$ signal, an alternative determination of the $\pi^+$ momentum is obtained from the RICH ring associated with the downstream track. Such RICH-based momentum is used to obtain an additional evaluation of the squared missing mass, called $M^2_\mathrm{miss(RICH)}$. The following condition is applied:
\begin{equation}
\label{eq:mmissCut2}
0 < M^{2}_{\mathrm{miss(RICH)}} <  0.07\text{~GeV}^{2}/c^4.
\end{equation}

The conditions applied reduce the contribution from events with a hard photon radiated in the final state.
The total number of selected events is $2 \times 10^{7}$ for downstream track momenta in the range 10--45~GeV/$c$. 
The number of tagged $\pi^0$ mesons is obtained, after the correction for the control-trigger downscaling factor, to be $N_{\pi^0} = 8 \times 10^9$ with an estimated background contamination below $10^{-5}$.
\section{Selection of the signal sample}
\label{sec:signalSelection}
The signal selection starts with the application to $\pi\nu\bar{\nu}$-trigger events of the same criteria (Section~\ref{sec:normSelection}) used to select the normalization sample.
Further conditions are applied to reject events with any in-time activity from $\pi^0 \to \gamma\gamma$ decays.
No signal must be found in the photon-veto detectors LAV, IRC, and SAC within 5~ns from the downstream track time.
In the LKr, synchronous energy deposits in nearby cells are grouped into clusters and no cluster must be present more than 10~cm away from the $\pi^+$ impact point in an energy-dependent time window; to recover residual inefficiencies of the clustering algorithm for soft photons, no energy deposit in the LKr above 0.5~GeV must be found in any $20 \times 20$~cm$^{2}$ area with centroid at least 30~cm away from the $\pi^+$ impact point.

Photons from $\pi^0$ decays can convert in the material before reaching the photon-veto detectors and the emitted electron-positron pair might escape detection in the photon vetoes.
This can occur because the $e^{\pm}$ energy is too low or because a shower is initiated, producing secondaries with small enough energy to escape detection. Moreover, if the conversion happens upstream of the STRAW spectrometer magnet, the particles can be bent into the regions between LAV stations or inside the beam tube, which are not instrumented. The most relevant contributions to the induced photon-veto inefficiency are due to conversions in the STRAW chamber material and in the RICH and CHOD mechanical structures.
Similar topologies can occur for $\pi^0$ Dalitz decays, $\pi^0 \to \gamma e^+ e^-$. 

Criteria making use of information from various detectors are used to reject the $e^+e^-$ backgrounds: 
no additional track segments compatible with the decay vertex must be reconstructed with the STRAW;
no positively charged track other than that from the charged pion must be found with the STRAW, with a CDA to the $\pi^+$ track below 30~mm and point of closest approach in the decay volume; no in-time signals must be found in the MUV0 and HASC detectors; no in-time signals must be found in the RICH except for those associated with the $\pi^+$ Cherenkov ring; fewer than four (one) in-time signals in the NA48-CHOD (CHOD) must be found, apart from those associated with the $\pi^+$.

To increase the veto capability against 
visible $\pi^0$ decays, an event is rejected if in-time signals are found in at least two out of three detectors NA48-CHOD, CHOD, and LKr that are spatially unrelated to the charged pion.

The signal selection achieves a rejection power of $\mathcal{O}(10^{8})$ against $\kppg$ with $\pi^0$ decays to visible particles for $\pi^+$ track momenta in the range 10--45~GeV/$c$ (see Section~\ref{sec:backgroundEvaluation}), corresponding to about 100 background events expected.
Taking the statistical and systematic uncertainties into account, the $\pi^+$ momentum interval 25--40~GeV$/c$ provides the expected upper limit with the lowest median and 68\%-coverage upper bound for $\mathrm{BR(\dec)}$ in the absence of signal.
\section{Selection and trigger efficiencies}
\label{sec:efficiencyCorrections}
%
The probability of wrongly vetoing events with neither photons nor photon conversions in the final state is evaluated with a sample of $K^+ \to \mu^+ \nu_{\mu}$ events selected in control-trigger data. 
To obtain this sample, $K^+$ decays to a single positively-charged particle are selected as described in Section~\ref{sec:normSelection}; a positive muon identification is obtained from the LKr, MUV1,2,3 and RICH detectors; the absolute value of the squared missing mass (under the muon mass hypothesis) is required to be below 0.005~GeV$^2$/$c^4$.
About $10^8$ $K^+ \to \mu^+ \nu_{\mu}$ decays are selected in the muon momentum range 10--45~GeV/$c$. 
The signal-selection conditions described in Section~\ref{sec:signalSelection} are then applied, resulting in a 54.2\% efficiency. The signal loss is dominated by accidental activity in the detector. Simulation studies indicate that the probability to satisfy the signal-selection conditions differs when considering muons or pions in the final state: about $2.5$\% (10\%) of the $K^+ \to \mu^+ \nu_{\mu}$ ($K^+\to\pi^+\pi^0$) decays are rejected by the signal-sample selection as a result of $\mu^+$ ($\pi^+$) interactions. 
The signal-selection efficiency is 
\begin{equation}
\label{eq:effisel}
\varepsilon_{\mathrm{sel}} = (50.1 \pm 1.6)\%,
\end{equation}
where the uncertainty quoted is systematic and accounts for the reliability of the simulation in reproducing the response to $\pi^+$ interaction secondaries.

The $\pi\nu\bar{\nu}$-trigger efficiency $\varepsilon_{\mathrm{trig}}$ is determined from control-trigger samples of $K^+ \to \pi^+ \pi^0$ and $K^+ \to \mu^+ \nu_{\mu}$ decays for the L0 and L1 $\pi\nu\bar{\nu}$-trigger levels, respectively, using the same techniques applied in the context of~\cite{RefNA62Pnn2017}. The trigger efficiency is evaluated given the selection conditions and depends
on the $\pi^+$ momentum: it varies from about 90\% at 10~GeV$/c$ to about 
75\% at 45~GeV$/c$, mainly due to the LKr-based L0 condition and to the STRAW-based L1 condition. A 3.5\% relative systematic uncertainty is estimated, due to the assumptions behind the technique used to evaluate offline the efficiency of the hardware conditions applied at L0. In the $\pi^+$ momentum range 25--40~GeV$/c$,
\begin{equation}
\label{eq:effitrig}
\varepsilon_{\mathrm{trig}} = (82.8 \pm 2.9)\%.
\end{equation}
\section{Evaluation of the expected background}
\label{sec:backgroundEvaluation}
Studies based on data and simulation show that, after signal selection, the residual background events are due to $\kppgIB$ decays where photons emitted in the 
$\pi^0 \to \gamma \gamma$ decay and in the $K^+$ decay via
the inner bremsstrahlung (IB) process,
are undetected. The IB amplitude is completely determined by the $K^+\to\pi^+\pi^0$ amplitude after QED corrections~\cite{RefLow}. 
The background from the direct-emission (DE) process $K^+\to\pi^+\pi^0\gamma$, including both the structure-dependent DE transition and the interference between the DE and IB amplitudes~\cite{RefRad1,RefRad2,RefRad3}, is negligible.
The total background from the decay channels $\kmng$, $\kppp$, $\kpeng$, $\kpmng$, $\kppeng$, $\kppmng$, and $\keng$ is expected to be below 0.1 events for the analyzed data sample, by extension of the background estimation for the NA62 search for $K^+ \to \pi^+ \nu \bar{\nu}$~\cite{RefNA62Pnn2017}. The background from $K^+ \to \pi^+ \nu \bar{\nu}$ is expected to be 0.2 events when assuming the Standard Model (SM) value for the BR~\cite{RefPNNSM}.

To evaluate the expected background, single-photon detection efficiencies are determined from an analysis of data control samples, as described in Section~\ref{sec:singlePhotonEfficiency}. 
The expected veto inefficiency $\eta_{\pi^0}$ for $\kppgIB$, $\pi^0\to\gamma\gamma$ decays is obtained from simulated events as the average probability that all photons escape detection, as described in Section~\ref{sec:expectedRejection}. The expected number of background events $N_{\mathrm{bkg}}$ is evaluated from the number of tagged $\pi^0$ mesons, $N_{\pi^0}$, by accounting for the $\pi\nu\bar{\nu}$-trigger efficiency, $\varepsilon_\mathrm{trig}$:
\begin{equation}
N_{\mathrm{bkg}} = N_{\pi^0} \times \eta_{\pi^0} \times \varepsilon_{\mathrm{trig}}.
\label{eq:back}
\end{equation}
\subsection{Single-photon detection efficiencies}
\label{sec:singlePhotonEfficiency}
The single-photon efficiency for each detector in the photon-veto system (LKr, LAV, IRC and SAC) is determined from a tag-and-probe (TP) analysis of control-trigger data.
Criteria in addition to those described in Section~\ref{sec:normSelection} allow selection of $K^+ \rightarrow \pi^+ \pi^0$, $\pi^0\to\gamma\gamma$ events with one or two photons from the $\pi^0$ decay reconstructed in the LKr.
Specific kinematic requirements are applied to reduce any contribution from the radiative process $K^+\to\pi^+\pi^0\gamma$, so that selecting one of the two photons from the $\pi^0$ decay as the tagging photon precisely determines the expected position and energy of the probed photon.

The energy and position of the tagging photon in the LKr, the 4-momenta of the $K^+$ and $\pi^+$ particles 
and the vertex position are used to determine the photon-veto detector towards which the probed photon is pointing (expected detector). An inaccurate reconstruction of the energy and/or position of the tagging photon may lead to an incorrect determination of the expected detector and/or 
the energy of the probed photon. To reduce the resulting systematic effects, events with in-time signals in the veto detectors other than the expected detector are rejected. The single-photon efficiency $\varepsilon_\mathrm{Data,TP}$ is defined as the fraction of remaining events with an in-time signal in the expected detector and is a function of the expected energy of the probed photon.
Because of the limited angular resolution of the probed photon direction, all the LAV stations are treated as a single detector in the efficiency evaluation; moreover, signals expected in the IRC and found in the SAC contribute to the IRC efficiency and \textit{vice versa}.
The statistical uncertainty in the efficiencies is evaluated with the Feldman-Cousins method~\cite{RefFCStatistics} for 68\% confidence intervals.

The efficiencies obtained must be corrected to account for two different effects: accidental activity alone can result in an in-time signal even in the absence of a photon-induced signal, while residual systematic effects can alter the result of the TP technique. 

The effect of accidental activity is independent of the photon energy, depends on the expected detector considered, and systematically leads to an efficiency higher than the true value.
For each expected detector, the probability $A$ that an in-time signal is detected due to accidental activity is evaluated with $K^+\to\mu^+\nu_{\mu}$ events selected from control-trigger data as described in Section~\ref{sec:efficiencyCorrections}. The results are:
\begin{equation}
\label{eq:acceptances}
A^\mathrm{LAV} = (35.3 \pm 1.1)\% \text{, } A^\mathrm{LKr} = (32.8 \pm 1.0)\%\text{, } A^\mathrm{IRC\mbox{-}SAC} = (26.4 \pm 0.8)\%,
\end{equation}
where the errors are dominated by systematic uncertainties, resulting from the same effects discussed for the selection efficiency in Equation~(\ref{eq:effisel}). The $A^{(i)}$ factors in Equation~(\ref{eq:acceptances}) refer to the probability of vetoing an event due to accidental activity \textit{only} in detector $i$.

Residual systematic effects inherent in the TP technique arise from incorrect determination of the energy and position of the tagging photon: these effects depend on both the expected energy and detector of the probed photon, and can lead to an efficiency higher or lower than the true value. As shown by MC simulation, if the tagging photon interacts before entering the sensitive volume of the LKr, the expected energy of the probed photon can be higher than the true value. The corresponding systematic uncertainties are determined with simulated $K^+\to\pi^+\pi^0$, $\pi^0\to\gamma\gamma$ decays. The true photon energy and the true detector to which the photon is pointing are given by the MC simulation and used to compute the true efficiency $\varepsilon_\mathrm{MC,true}$; the TP method is applied after simulation of the detector response and the resulting single-photon efficiencies $\varepsilon_\mathrm{MC,TP}$ are computed as for the data. The ratio
\begin{equation}
\label{eq:bias}
b = \frac{1-\varepsilon_\mathrm{MC,TP}}{1-\varepsilon_\mathrm{MC,true}}
\end{equation}
is measured separately for the LAV and LKr detectors and taken as an evaluation of the bias induced by the TP method. Extremely large statistics would be needed to evaluate the bias related to the IRC and SAC detectors and therefore their bias ratio is assumed to be 1 in the following.
As demonstrated by the method validation (Section~\ref{sec:validation}), the systematic uncertainty due to the IRC-SAC bias is expected to be a sub-leading effect after the optimization of the signal selection. 

\begin{figure}[htb]
\center
\begin{minipage}{0.496\textwidth}
\includegraphics[width=1.0\textwidth]{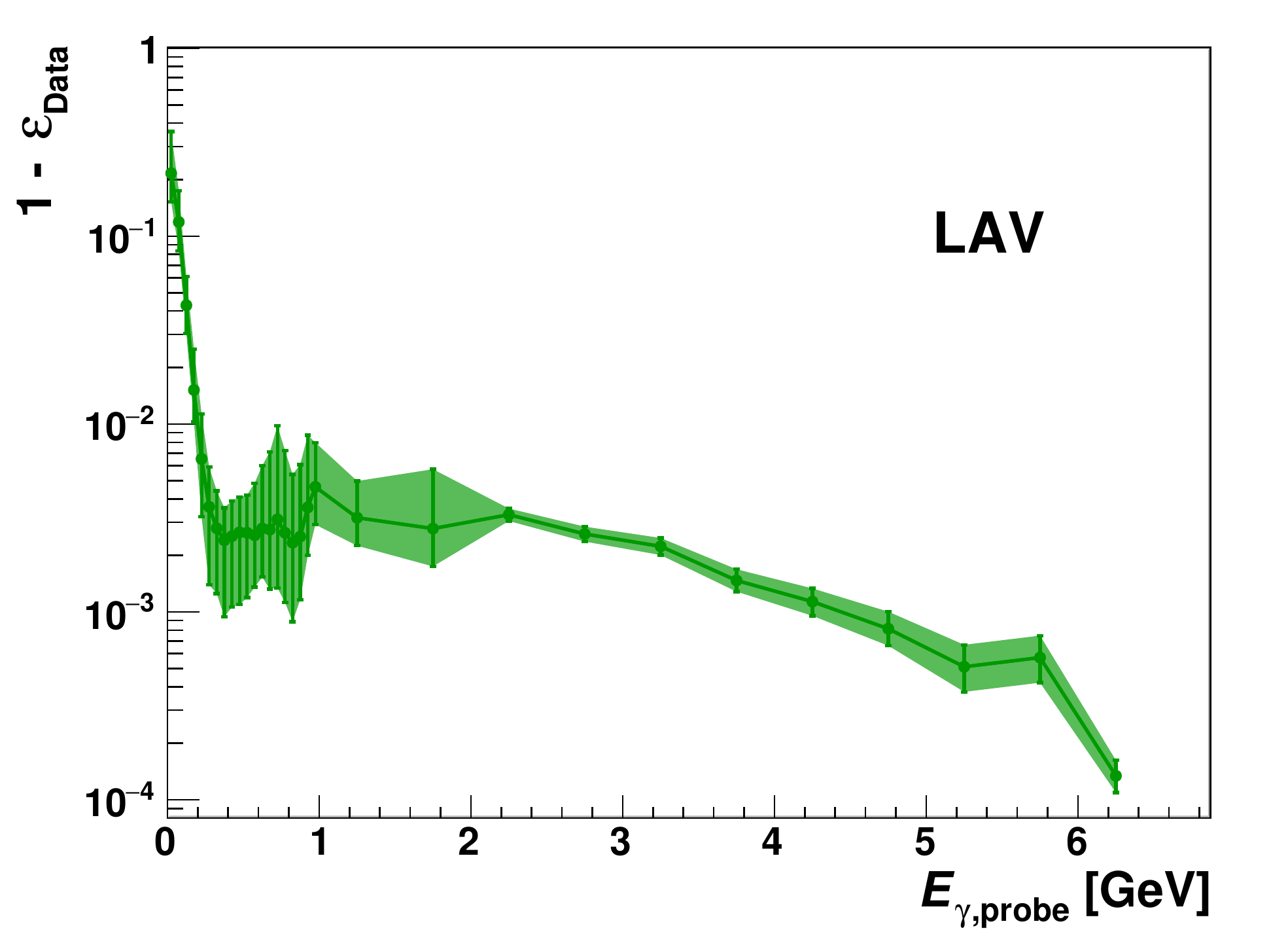}
\end{minipage}
\begin{minipage}{0.496\textwidth}
\includegraphics[width=1.0\textwidth]{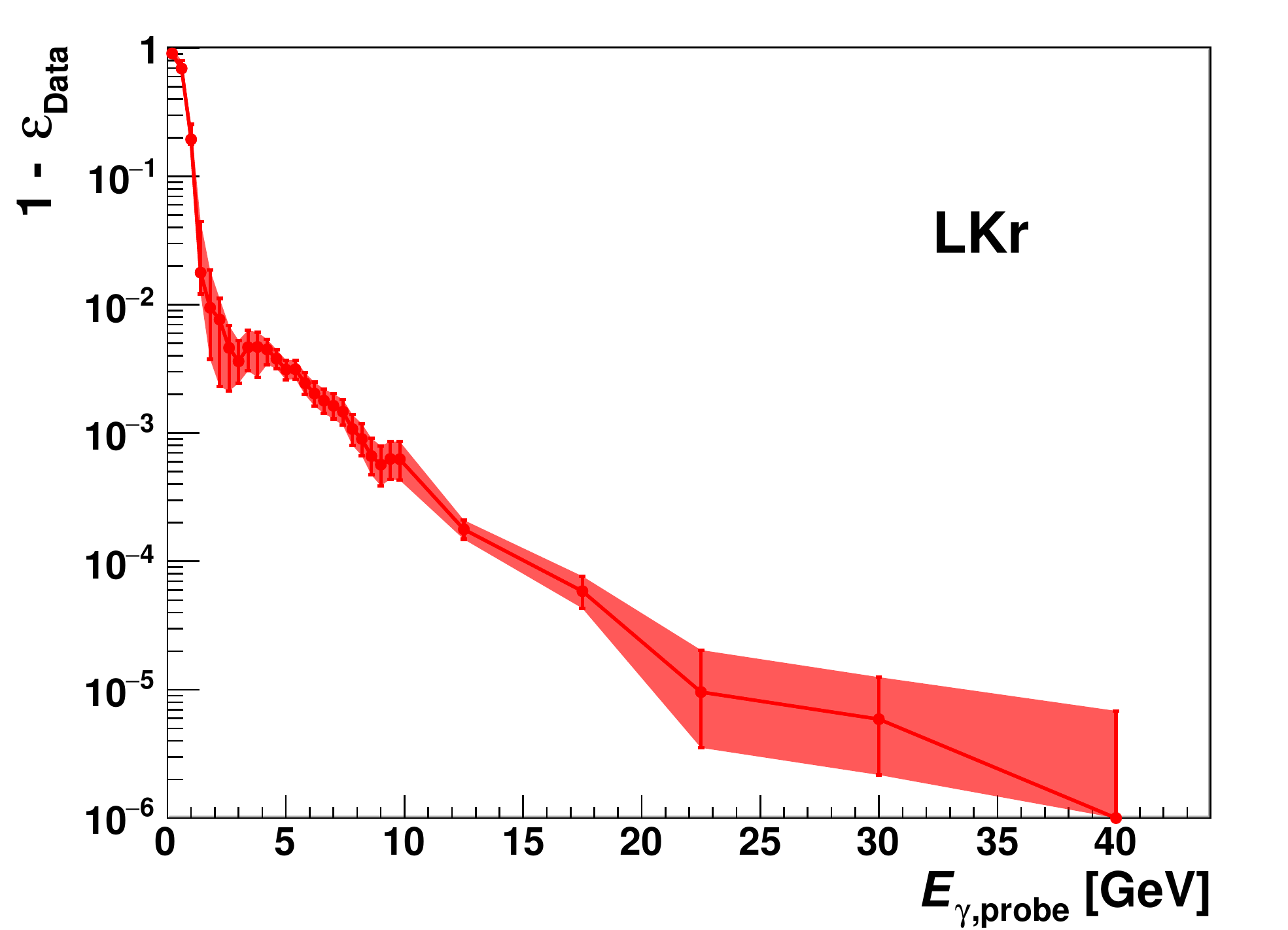}
\end{minipage}
\begin{minipage}{0.496\textwidth}
\includegraphics[width=1.0\textwidth]{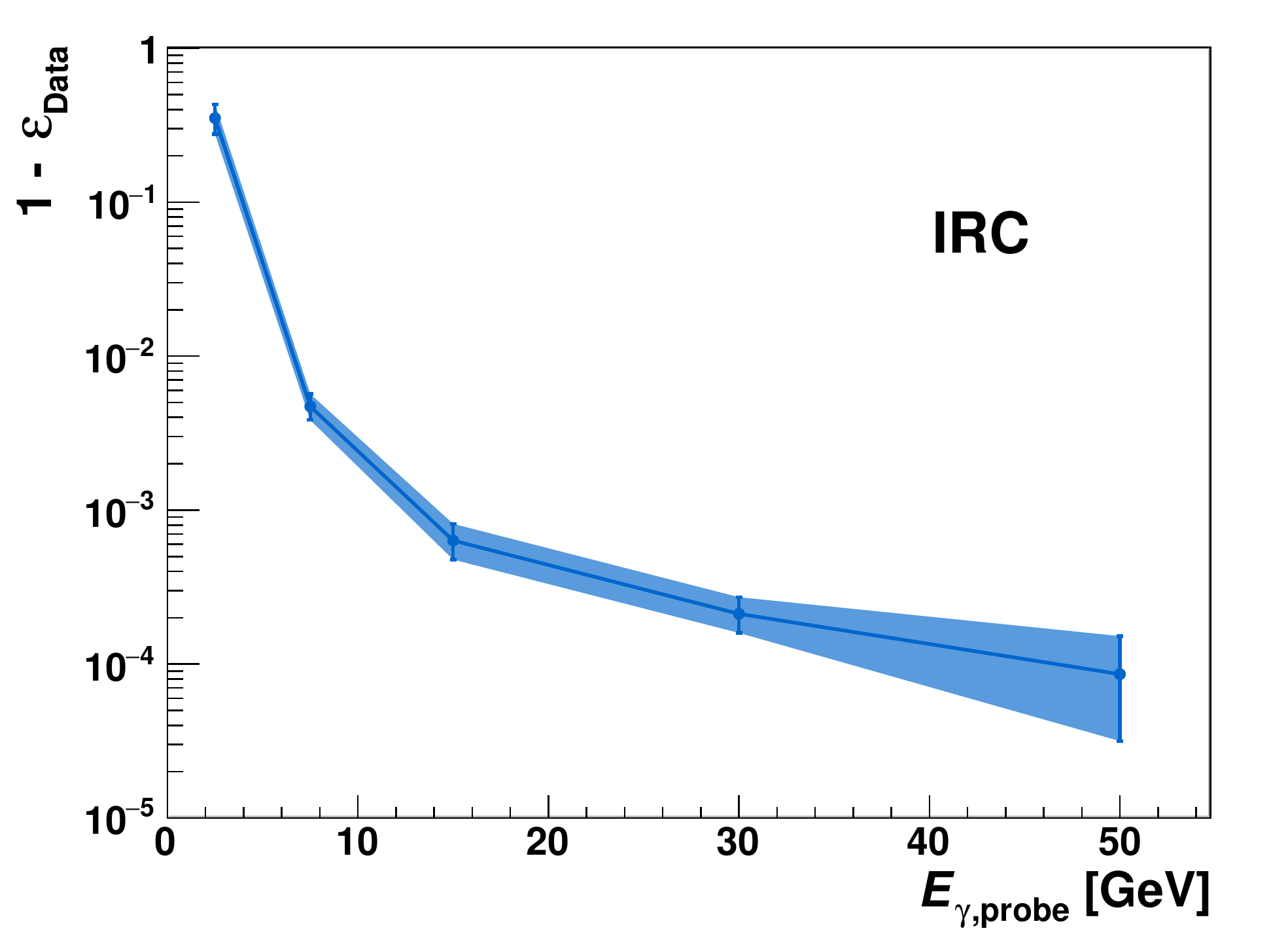}
\end{minipage}
\begin{minipage}{0.496\textwidth}
\includegraphics[width=1.0\textwidth]{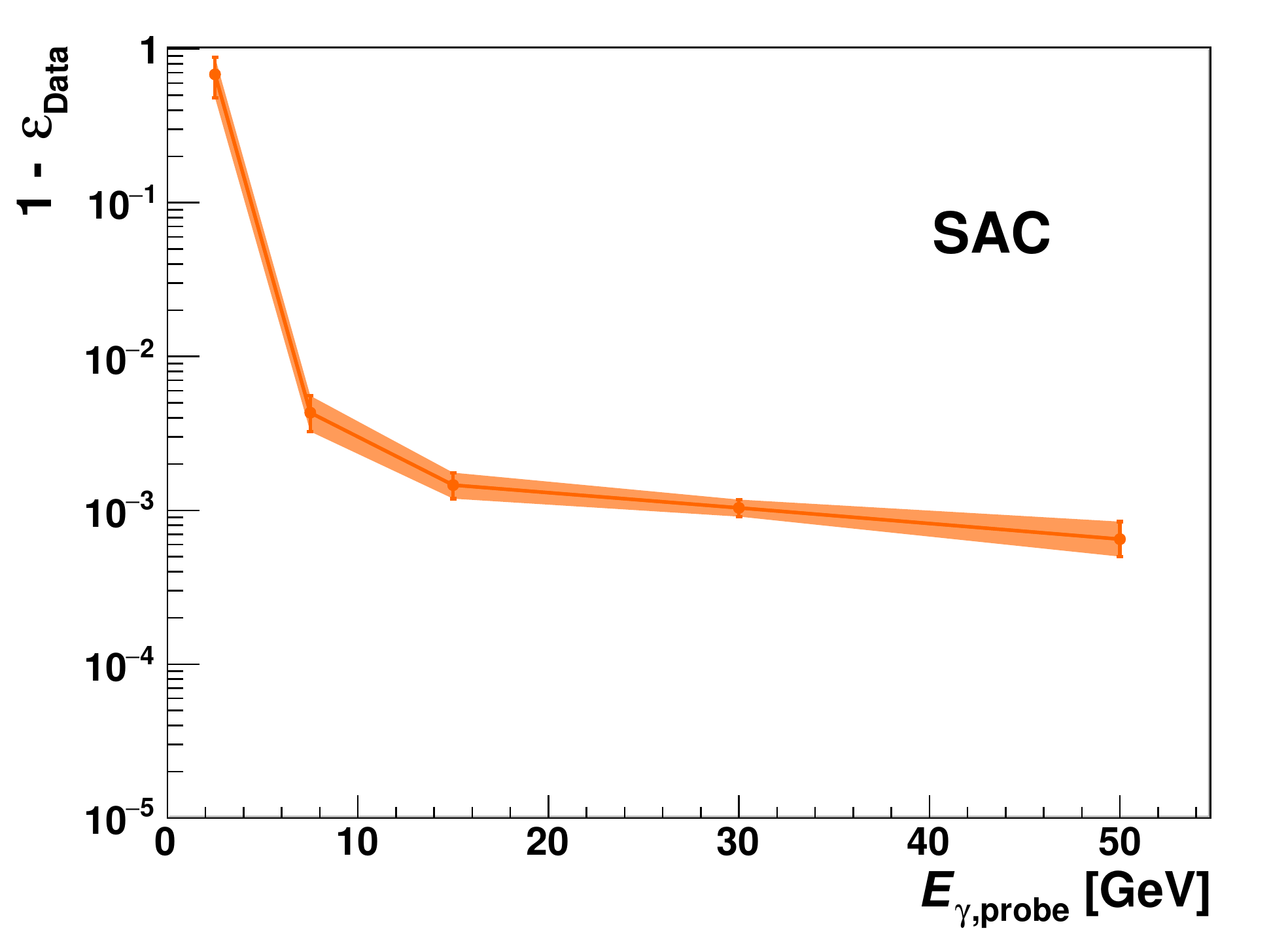}
\end{minipage}
\caption{Detection inefficiency as a function of the energy of the probed photon $E_{\gamma,\mathrm{probe}}$, for the LAV (top left), LKr (top right), IRC (bottom left) and SAC (bottom right) detectors.}
\label{fig:singlePhotonEfficiency}
\end{figure}

The single-photon inefficiencies used for the background evaluation, $1 - \varepsilon_\mathrm{Data}^{(i)}$, are obtained from the inefficiencies from the TP method, $1 - \varepsilon_\mathrm{Data,TP}^{(i)}$, as follows:
\begin{equation}
\label{eq:singlepe}
1-\varepsilon_\mathrm{Data}^{(i)} = \frac{1-\varepsilon_\mathrm{Data,TP}^{(i)}}{\left(1-A^{(i)}\right)b^{(i)}} ,
\end{equation}
where the index $i$ refers to each photon-veto detector (LAV, LKr, IRC-SAC), and the quantities $A^{(i)}$ and $b^{(i)}$ are given by Equations~(\ref{eq:acceptances}) and~(\ref{eq:bias}).
The inefficiencies are shown as a function of the photon energy in Figure~\ref{fig:singlePhotonEfficiency}, where only the statistical uncertainties are displayed. The systematic uncertainties in the expected background rejection are discussed in Section~\ref{sec:expectedRejection}.

\subsection{Expected veto inefficiency for \boldmath{$\kppgIB$} decays}
\label{sec:expectedRejection}
A simulation of the decay chain $\kppgIB$, $\pi^0\to\gamma\gamma$ is used to evaluate the expected veto inefficiency. For each of the photons from the $K^+$ and $\pi^0$ decays, the inefficiencies of Equation~(\ref{eq:singlepe}) are combined to obtain an event weight equal to the probability that all the photons escape detection. The expected veto inefficiency, $\eta_{\pi^0}$, is obtained from the average of the event weights.

\begin{figure}[htb]
\center
\includegraphics[width=0.7\textwidth]{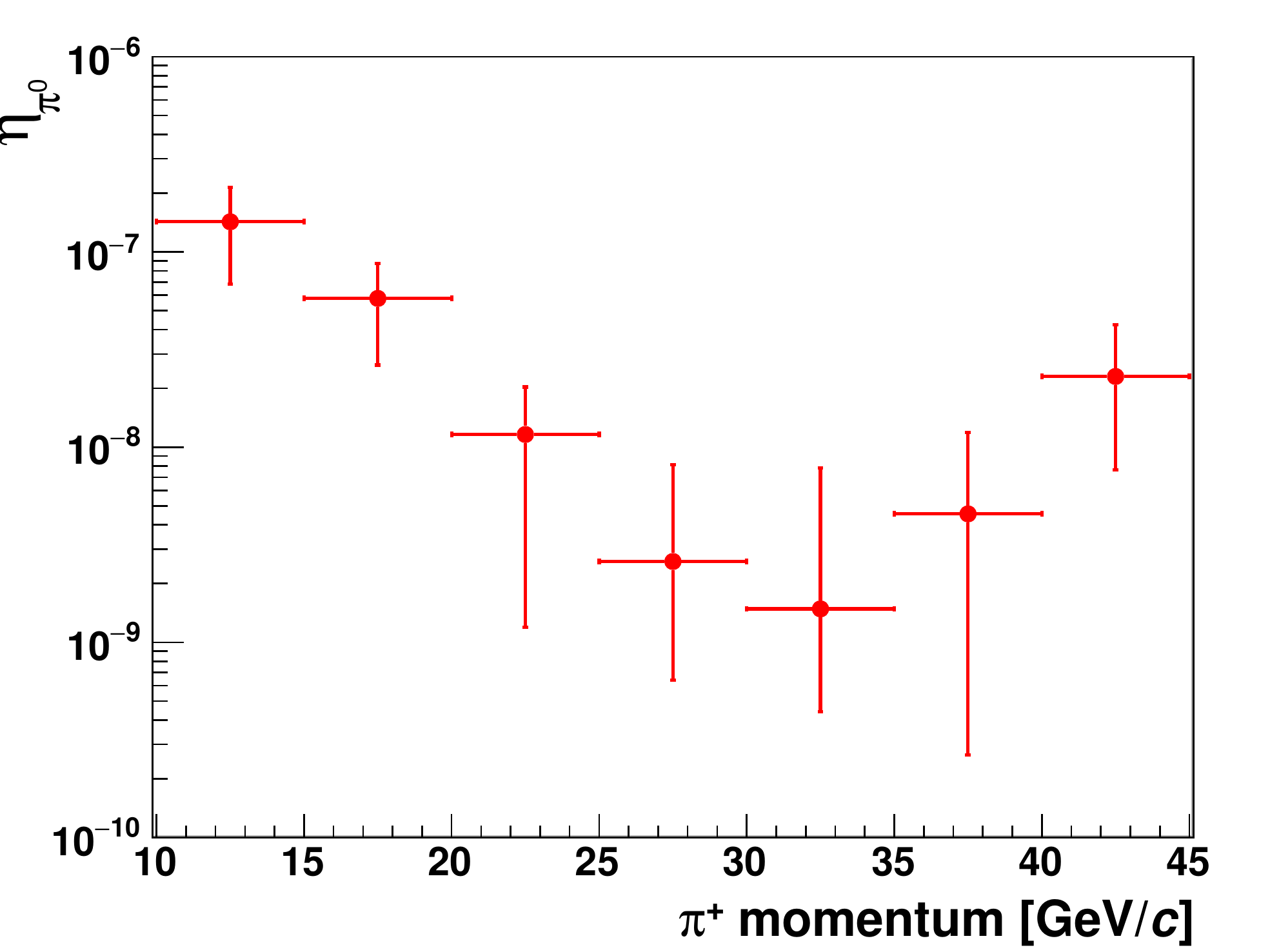}
\caption{Expected veto inefficiency $\eta_{\pi^0}$ as a function of the charged pion momentum.}
\label{fig:expectedRejection}
\end{figure}

The result strongly depends on the $\pi^+$ momentum, as shown in Figure~\ref{fig:expectedRejection}: a $\pi^+$ momentum above 40 GeV$/c$ corresponds to a $\pi^0$ momentum below 35~GeV$/c$ and to a higher probability that one of the photons escapes detection by passing through one of the regions not instrumented between two LAV stations; a $\pi^+$ momentum below 15--20~GeV$/c$ corresponds to a higher probability that one of the two $\pi^0$-daughter photons traverses the beam pipe at a grazing angle and converts to an $e^+e^-$ pair escaping detection. 
The expected veto inefficiency strongly depends on the momentum of the radiative photon from the IB process: decays with a radiative photon harder than 10~keV in the $K^+$ rest frame are rejected with an inefficiency at the level of or below $10^{-10}$. 
In the $\pi^+$ momentum range 25--40~GeV$/c$,
\begin{equation}
\label{eq:epsp0}
\eta_{\pi^0} = \left( 2.8^{+5.9}_{-2.1} \right)\times 10^{-9}.
\end{equation}

Three sources of uncertainty are considered:
\begin{itemize}
    \item the limited statistics of the MC simulation sample. The resulting uncertainty is evaluated as the standard deviation of the values of $\eta_{\pi^0}$ obtained from 20 MC sub-samples;
    \item the statistical uncertainties on the underlying inefficiencies $\varepsilon_\mathrm{Data}^{(i)}$. The resulting uncertainty is evaluated from the distribution of the values of $\eta_{\pi^0}$ obtained by varying each inefficiency $\varepsilon_\mathrm{Data}^{(i)}$ within its asymmetric uncertainty. In particular, each point
in energy $\varepsilon^{(i)}_\mathrm{Data}$ is varied independently within its uncertainty providing a set of 1\,000 alternative inefficiency curves for the four detectors. A 68.8\%-coverage band is quoted and asymmetric errors are evaluated with respect to the value of $\eta_{\pi^0}$ obtained from the central values of $\varepsilon_\mathrm{Data}^{(i)}$;
    \item the systematic uncertainty of the correction applied for the TP bias in Equation~(\ref{eq:bias}). 
    The direction and energy of the probed photon are largely determined from the energy and position measurements of the tagging photon in the LKr, so that the LKr becomes the most significant source of systematic effects. 
    In particular, the reliability of the MC simulation in describing the non-Gaussian tails of the LKr energy resolution is studied with $\kppg$ decays with $\pi^0\to e^+e^-\gamma$. 
    The reconstructed charged particle momenta are used to determine the expected 4-momentum of the photon from the $\pi^0$, which is required to point to the sensitive volume of the LKr.
    The presence of a photon signal in the LKr within 50~cm of the expected impact point is required.
    The distribution of the difference between the expected and reconstructed photon energy from data and MC simulation are compared and the distribution from data is found to have longer tails. The observed data-MC difference is due to the simulation of the LKr response, while the reconstruction of charged particle momenta has a negligible effect. The resolution observed in data is parametrized as a function of the photon energy and used to simulate the LKr detector response to obtain an alternative evaluation of $\varepsilon_\mathrm{MC,TP}$. 
    A systematic uncertainty is assigned as the change in the result obtained for $\eta_{\pi^0}$ when using the alternative evaluation of $\varepsilon_\mathrm{MC,TP}$ accounting for the tails in the LKr energy response. The LAV inefficiency is the most sensitive to the LKr energy resolution model and the corresponding systematic uncertainty in the region $E_\mathrm{\gamma,probe}>1$~GeV is singled out in Table~\ref{tab:systematics}.
\end{itemize}
The second source of uncertainty dominates the total error. In particular, the uncertainty on the three bins with highest energy of the LKr inefficiency curve is the leading contributor to the overall error on $\eta_{\pi^0}$: these events correspond to $\pi^0\to\gamma\gamma$ decays with one photon in the LKr and the other emitted at such large angle that it escapes detection passing between two LAV stations.
Additional sources of systematic uncertainty due to accidental in-time activity in the photon-veto detectors and trigger effects are found to be negligible. Further sources of systematic uncertainty specific to the identification of in-time signals (Section~\ref{sec:signalSelection}) are considered. In particular, the criteria defining the regions of NA48-CHOD, CHOD, and LKr spatially unrelated to the charged pion are 
varied, leading to negligible changes of the results.

The expected background is evaluated from Equation~(\ref{eq:back}) and the results are listed in Table~\ref{tab:background}. It must be noted that the uncertainties on $N_\mathrm{bkg}$ for the
momentum bins in the range 20--40~GeV/$c$ are highly correlated. The various sources of uncertainties on the expected veto inefficiency $\eta_{\pi^0}$ in the $\pi^+$ momentum region 25--40~GeV/$c$ are summarized in Table~\ref{tab:systematics}.

\begin{table}[htb]
\center
\caption{Summary of the number of tagged $\pi^0$ mesons ($N_{\pi^0}$), expected veto inefficiency ($\eta_{\pi^0}$), $\pi\nu\bar{\nu}$-trigger efficiency ($\varepsilon_\mathrm{trig}$), and expected number of background events ($N_\mathrm{bkg}$) in bins of the $\pi^+$ momentum.}
\label{tab:background}
\vspace{5pt}
\begin{tabular}{c|d{4.0}|c|c|c}
\toprule
$\pi^+$ Momentum~[GeV/$c$] & \multicolumn{1}{c|}{$N_{\pi^{0}}$~[$10^6$]} & $\eta_{\pi^0}$~[$10^{-8}$] & $\varepsilon_\mathrm{trig}$~[\%] & $N_{\mathrm{bkg}}$ \\
\midrule
10 -- 15 & 84 & $14.3^{+7.1}_{-7.4}$ & $91.6 \pm 3.2$ & $11.1^{+5.5}_{-5.8}$ \\ [0.65ex]
15 -- 20 & 923 & $~5.8^{+3.0}_{-3.2}$  & $86.3 \pm 3.0$ & $~~46^{+24}_{-25}$ \\ [0.65ex]
20 -- 25 & 1\,383 & $~1.16^{+0.87}_{-1.04}$ & $85.2 \pm 3.0$ & $~~14^{+10}_{-12}$ \\ [0.65ex]
25 -- 30 & 1\,612 & $~0.26^{+0.55}_{-0.20}$ & $84.3 \pm 3.0$ & $~~3.5^{+7.5}_{-2.7}$ \\ [0.65ex]
30 -- 35 & 1\,504 & $~0.15^{+0.64}_{-0.10}$ & $83.1 \pm 2.9$ & $~~1.9^{+7.9}_{-1.3}$ \\ [0.65ex]
35 -- 40 & 1\,313 & $~0.45^{+0.74}_{-0.43}$ & $80.7 \pm 2.8$ & $~~4.8^{+7.8}_{-4.5}$ \\ [0.65ex]
40 -- 45 & 1\,058 & $~2.3^{+1.9}_{-1.5}$ & $75.5 \pm 2.6$ & $~~18^{+15}_{-12}$ \\ [0.65ex]
\bottomrule
\end{tabular}
\end{table}

\begin{table}[htb]
\center
\caption{Summary of the uncertainties ($\Delta$) of the single-photon inefficiencies $1-\varepsilon_\mathrm{Data}$ and of the veto inefficiency $\eta_{\pi^0}$ in the $\pi^+$ momentum region 25--40~GeV/$c$. 
}
\label{tab:systematics}
\vspace{5pt}
\begin{tabular}{c|c|c}
\toprule
Source & $\Delta(1 - \varepsilon_{\mathrm{Data}})$ & $\Delta(\eta_{\pi^0})$ \\
\midrule
Uncertainty on $\varepsilon_\mathrm{Data}$ & See Figure~\ref{fig:singlePhotonEfficiency} & $~~~^{+5.9}_{-2.1} \times 10^{-9}$  \\ [0.5ex]
Data driven TP bias, LAV & $~~^{+0.5}_{-0.2}\times10^{-2}$, $E_\mathrm{\gamma,probe}>~1$~GeV & $~~~^{+0.6}_{-0.3} \times 10^{-9}$ \\ [0.5ex]
Cut variation, in-time activity, trigger effects &  & negligible \\ [0.5ex]
\midrule
Total &  &  $~~~^{+5.9}_{-2.1} \times 10^{-9}$ \\
\bottomrule  
\end{tabular}
\end{table}

\section{Analysis validation}
\label{sec:validation}
The reliability of the evaluation of $\eta_{\pi^0}$ is verified with data in the sideband regions of the $\pi^+$ momentum, $p_{\pi^+}$, in which $\eta_{\pi^0}$ is large enough that the current experimental limit on $\dec$~\cite{RefBNLE949Paper} is sufficient 
for excluding the presence of signal events. 

The low-momentum sideband region consists of events with $10\text{~GeV/}c < p_{\pi^+} < 15\text{~GeV/}c$, for which
the expected veto inefficiency is of the order of $10^{-7}$ and any $\dec$ signal can be excluded at 90\% CL by the current experimental limit. 
Using the CL$_\mathrm{s}$ method~\cite{RefCLsPaper}, frequentist 90\% confidence intervals in the absence of signal are determined for the number of expected background events and, after simulation of the number of observed events, for the number of background-subtracted signal events.
No excess is observed in data: the number of observed events is compatible with statistical fluctuations of the background only. In the low-momentum sideband region, the single-photon inefficiencies of the IRC and SAC detectors for photon energies above 10~GeV are the leading contributors to the total background rejection. The test performed 
validates the uncertainty for the IRC-SAC inefficiency. A bias on the SAC (IRC) inefficiency for photon energies above 10~GeV exceeding $\pm5\times10^{-4}$ ($^{+2.0}_{-1.0}\times10^{-4}$) would have produced a deviation of $\pm1\sigma$ between expected and observed values of $\eta_{\pi^0}$. The bias quoted approximately corresponds to two (three) times the total uncertainty on the SAC (IRC) inefficiency and to an uncertainty on $\eta_{\pi^0}$ of $\pm0.1\times10^{-9}$ ($^{+0.5}_{-0.2}\times10^{-9}$).

For $\pi^+$ momenta above 15~GeV$/c$, the LKr inefficiency dominates the expected background rejection, with a sub-leading contribution from the LAV. Given an expected veto inefficiency of the order of $10^{-9}$--$10^{-8}$, the current experimental limit~\cite{RefBNLE949Paper} does not provide a useful constraint, and a different test is performed. 
The background rejection inefficiency is artificially increased to the level of $10^{-7}$--$10^{-6}$ by the introduction of an artificial energy-independent contribution $\Delta\varepsilon$ to the single-photon LAV inefficiency. 
This procedure has two advantages: i) The current experimental limit excludes the presence of signal events for a rejection power of $10^{7}$ or less, and the background expectation can be directly validated by comparison to data, maintaining blind analysis principles; ii) Events with one photon from the $\pi^0$ in the LAV and the other in the LKr become dominant, allowing a validation of the uncertainty on the LKr inefficiency.
For each value of $\Delta\varepsilon$ considered, the expected rejection power is validated against data, where, in a fraction of events equal to $\Delta\varepsilon$, no use is made of information from the LAV. To avoid any correlation with the signal sample, the procedure is applied to control-trigger data after the selection of the normalization sample (Section~\ref{sec:normSelection}).
The values of the expected and observed rejection power are compatible within the quoted uncertainties. 
The result 
validates the uncertainty for the LKr inefficiency. 
A bias on the LKr inefficiency for photon energies above 20~GeV exceeding $^{+4.6}_{-1.0}\times10^{-6}$ would have produced a deviation of $\pm1\sigma$ between expected and observed values of $\eta_{\pi^0}$. The bias quoted approximately corresponds to 1.5 times the total uncertainty on the LKr inefficiency and to an uncertainty on $\eta_{\pi^0}$ of $^{+4.6}_{-1.0}\times10^{-9}$.

\section{Results}
\label{sec:ulEval}
In the $\pi^+$ momentum range 25--40~GeV$/c$, $4.4 \times 10^{9}$ $\pi^0$ mesons are tagged.
With the veto inefficiency evaluated from Equation~(\ref{eq:epsp0}) and the trigger efficiency evaluated from Equation~(\ref{eq:effitrig}), the expected number of backgrounds events from Equation~(\ref{eq:back}) is $N_\mathrm{bkg}=10^{+22}_{-8}$.
In the absence of a signal, the 90\% CL upper limit on the number of signal events $N_\mathrm{s}$ is expected from the CL$_\mathrm{s}$ technique to have a median of 7.2 and a 68\% (95\%) coverage interval of 4.3--12.0 (2.4--17.7).

\begin{figure} [htb]
\centering
\begin{minipage}{0.496\linewidth}
\includegraphics[width=1.\linewidth]{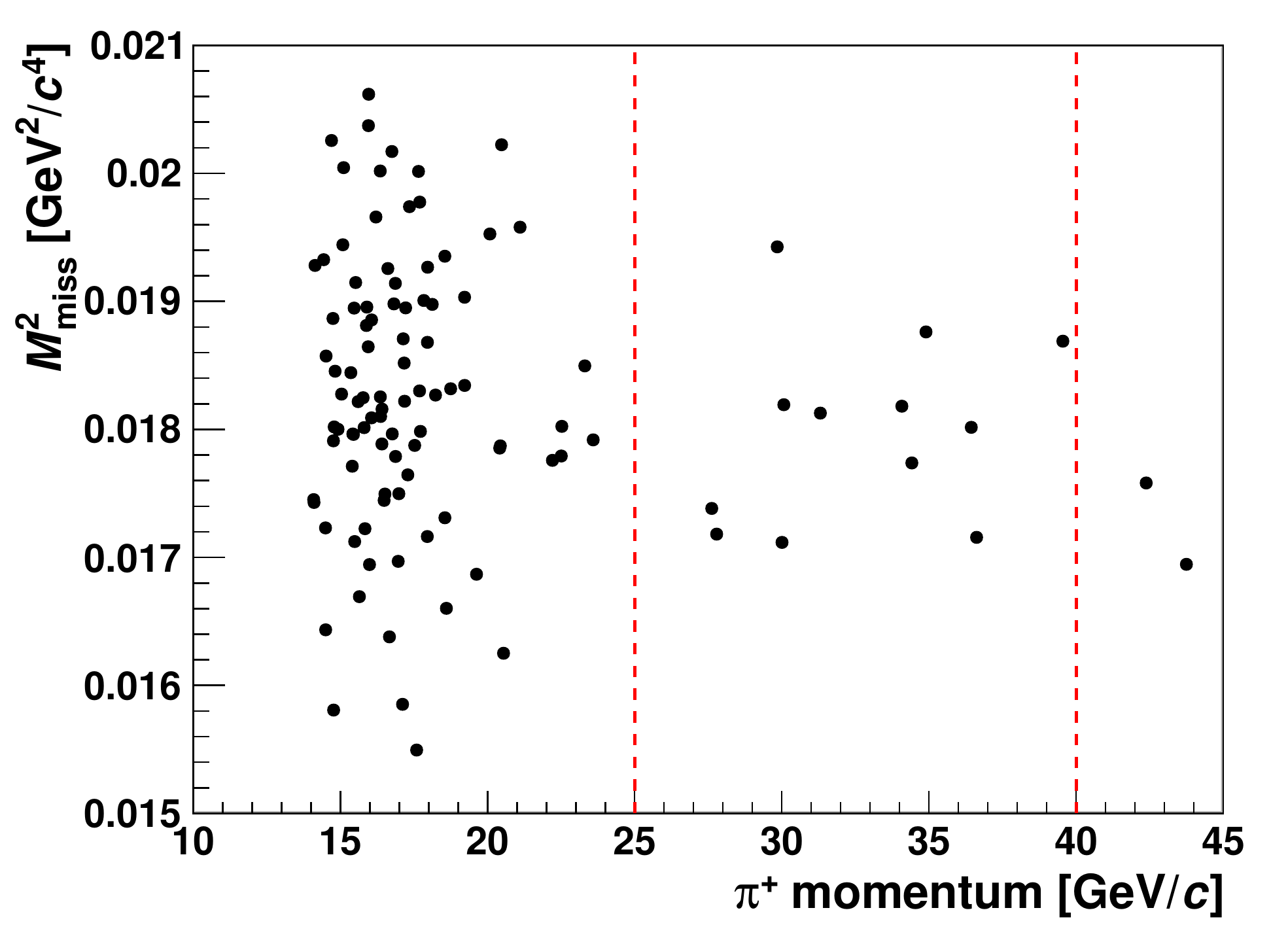}
\end{minipage}
\begin{minipage}{0.496\linewidth}
\includegraphics[width=1.\linewidth]{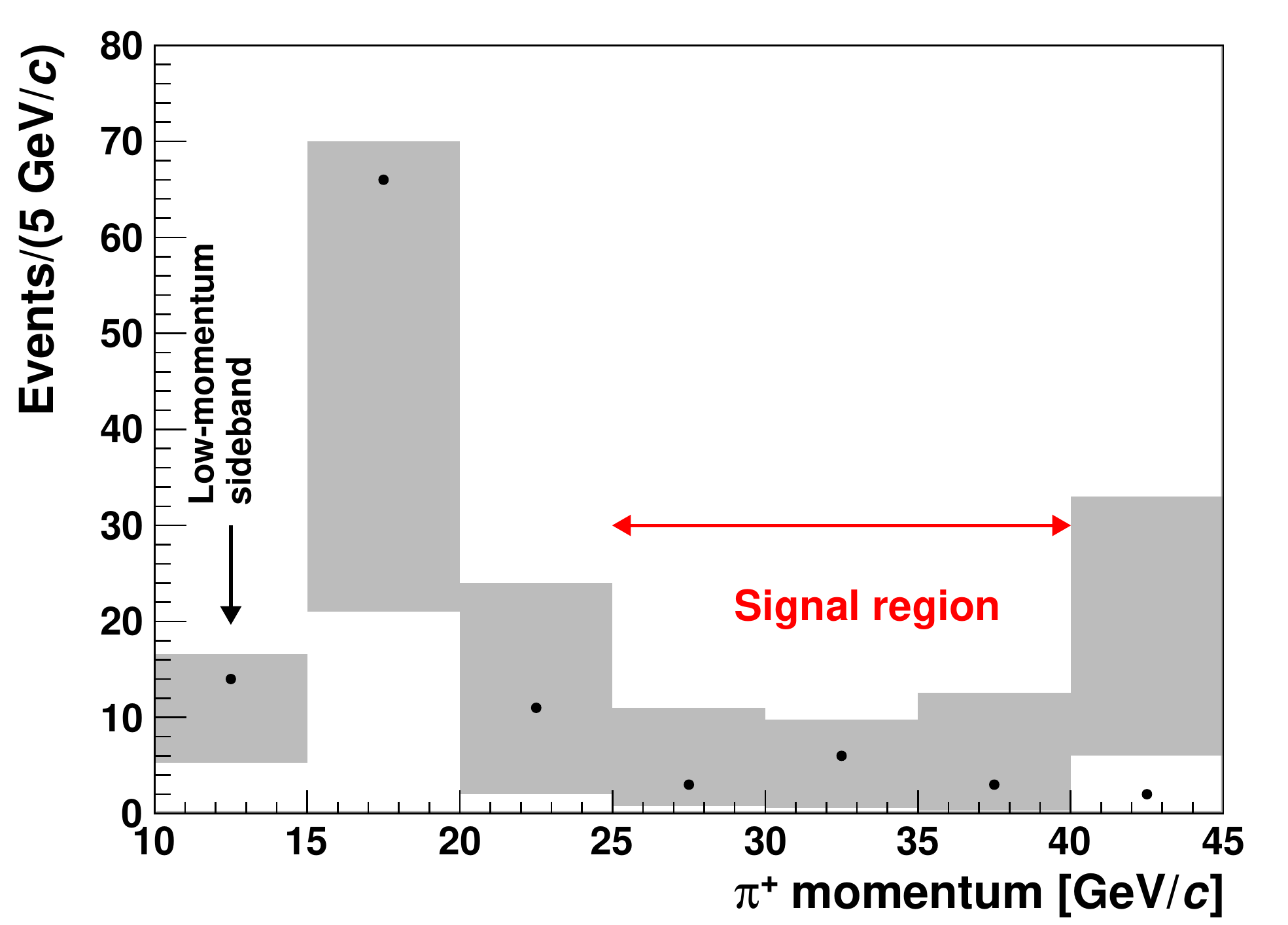}
\end{minipage}
\caption[]{Distributions for $\pi\nu\bar{\nu}$-trigger events selected with the criteria described in Section~\ref{sec:signalSelection} apart from the cut on the $\pi^+$ momentum. Left: squared missing mass of Equation~(\ref{eq:mmiss}) as a function of the $\pi^+$ momentum. Right: $\pi^+$ momentum for data (dots) and expected background (filled areas).}
\label{fig:SignalSample}
\end{figure}

In the sample of $\pi\nu\bar{\nu}$-trigger data selected with the criteria described in Section~\ref{sec:signalSelection}, 12 events consistent with a $\dec$ decay
are observed. 
The number of observed counts is compatible within the uncertainties with a pure-background hypothesis.
The CL$_\mathrm{s}$ method is applied to determine a frequentist 90\% confidence upper limit on the number of signal events: $N_{\mathrm{s}} < 8.24$. From Equation~(\ref{eq:BrPi0Invisible}), the upper limit on the branching ratio for $\dec$ is
\begin{equation}
\mathrm{BR} (\dec) < 4.4 \times 10^{-9}.
\label{eq:BRobserved}
\end{equation}
This result improves on the previously most stringent upper limit~\cite{RefBNLE949Paper} by a factor of about 60.
A light, feebly-coupled, spin-1 boson $U$ beyond the SM might generate the decay channel $\pi^0\to UU$. The $U$ boson might be invisible, decaying to a dark matter particle pair or to a $\nu\overline{\nu}$ pair. The vector couplings of the $U$ boson to quarks can be constrained by the present result, allowing an improvement by a factor of 2.7~\cite{Fayet2006}.

As an illustration of the agreement of the observed data with the expected background, a study is performed based on $\pi\nu\bar{\nu}$-trigger events selected with the criteria described in Section~\ref{sec:signalSelection} apart from the cut on the $\pi^+$ momentum. The distribution of the squared missing mass $M^{2}_\mathrm{miss}$ of Equation~(\ref{eq:mmiss}) is shown as a function of the $\pi^+$ momentum in the left panel of Figure~\ref{fig:SignalSample}. The $M^{2}_\mathrm{miss}$ distribution for the 12 events in the signal region, enclosed within the vertical dashed lines, is compatible with that expected from $K^+\to\pi^+\pi^0(\gamma)$ decays. The $\pi^+$ momentum distribution for data is shown along with the expected background in the right panel of Figure~\ref{fig:SignalSample}: data and background are compatible, with a combined probability in the full momentum range of 13\%.
%
%
\subsection{Interpretation in terms of a \boldmath{$K^+\to\pi^+X$} decay}
\label{sec:Axion}
\begin{figure}[htb]
\center
\includegraphics[width=0.7\textwidth]{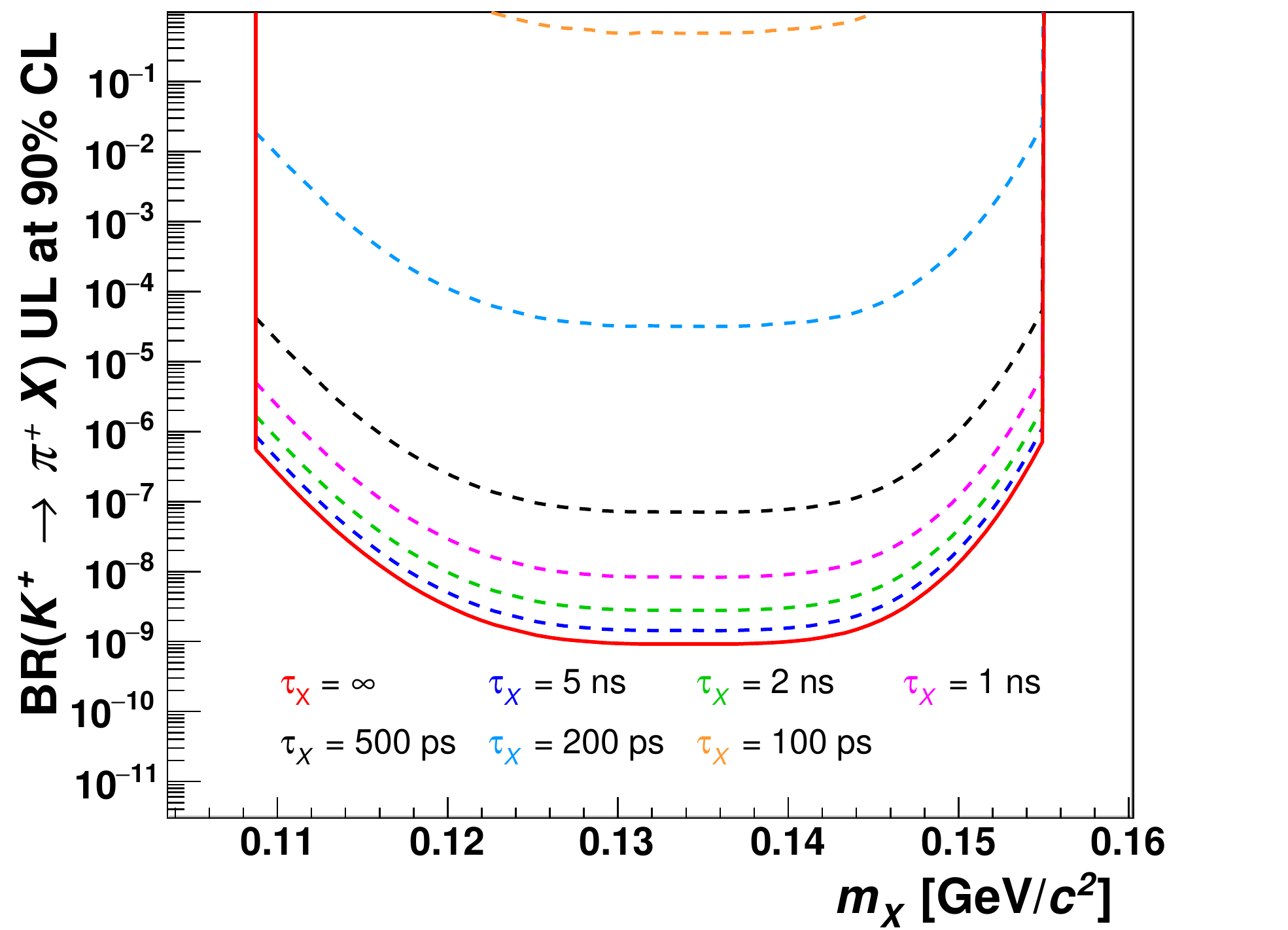}
\caption{The upper limit on the branching ratio for a decay $K^+\to\pi^+X$ is indicated by the solid line, where $X$ is a particle (or a system of particles) with mass $m_X$ escaping detection. 
The dashed lines are obtained with the assumption that $X$ is a single particle with values of the lifetime, $\tau_X$, as shown in the figure.
}
\label{fig:axionBranching}
\end{figure}

With no modifications to the analysis, the result can be converted into a limit for the decay $K^+\to\pi^+X$, where $X$ stands for any system with mass $m_X$ assumed to escape detection because it is long-lived and interacts feebly with SM particles. The kinematic condition of Equation~(\ref{eq:mmissCut}) restricts the search to a range of $m_X$ of approximately 0.110--0.155~GeV$/c^2$. From Equation~(\ref{eq:BrPi0Invisible}), the branching ratio for $K^+\to\pi^+X$ is
\begin{equation}
\label{eq:BrALP}
\mathrm{BR}(K^+\to\pi^+ X) = \frac{N_{\mathrm{s}}}{N_{K^+} \times R(m_{X}) \times \varepsilon_{\mathrm{sel}} \times \varepsilon_{\mathrm{trig}}} \text{ ,}
\end{equation}
where the number of $K^+$ decays in the sample ($N_{K^+}$) is obtained as 
\begin{equation}
N_{K^+} = \frac{N_{\pi^0}}{\mathrm{BR}(\kppg)\times\mathrm{BR}(\pi^0\to\gamma\gamma)}    
\end{equation}
and $ R(m_X)$ is the acceptance of the kinematic conditions of Equations~(\ref{eq:mmissCut}) and~(\ref{eq:mmissCut2}) for $K^+\to\pi^+X$ normalized to that for $K^+\to\pi^+\pi^0$, where radiative effects are assumed to cancel out. The resulting 90\% CL model-independent upper limit is shown as a solid line in Figure~\ref{fig:axionBranching}.
Assuming that $X$ is a single particle dominantly decaying to SM particles, different hypotheses have been made for the value of the $X$ lifetime, $\tau_{X}$. Whenever the $X$ particle decays inside the sensitive volume, the event is conservatively assumed to be rejected by the analysis veto conditions. The corresponding upper limits are shown as dashed lines: the lower the value of $\tau_{X}$, the higher the probability that the $X$ decay products satisfy the conditions to veto the event and, correspondingly, the weaker the upper limit. 

The present result is relevant to searches for axion-like particles (ALP) with dominant fermion coupling and to searches for dark scalars, as discussed in the next two sections.
%
\subsection{Interpretation in terms of ALP production from \boldmath{$K^+$} decays}
The ALP model denoted as \textit{BC10} in~\cite{RefPBC_Document} assumes an ALP field $a$ interacting with SM fermions according to the Lagrangian:
\begin{equation}
\mathcal{L}_{\mathrm{SM}} = \frac{\partial_{\mu} a}{f_{\ell}} \sum_{\alpha} \bar{\ell_{\alpha}} \gamma_{\mu} \gamma_{5} \ell_{\alpha} +  \frac{\partial_{\mu} a}{f_{q}} \sum_{\beta} \bar{q_\beta} \gamma_{\mu} \gamma_{5} q_{\beta},
\label{eq:AxionPBCLagrangian}
\end{equation}
where, for simplicity, the coupling parameters $f$ are assumed to be universal for quark ($q$) and lepton ($\ell$) fields: $f_{q} = f_{\ell}$. In \textit{Yukawa-like} scenarios, the interactions between the ALP $a$ and the charged SM fermions arise from a mixing with the Higgs boson, so that the couplings are proportional to the fermion masses. The corresponding Yukawa coupling is defined as $g_{Y} = \frac{2v}{f_{q}}$, where $v \simeq 246$~GeV is the vacuum expectation value of the SM Higgs field. The dynamics of Equation~(\ref{eq:AxionPBCLagrangian}) induces a flavour-changing neutral current transition $K^+ \to \pi^+ a$.

\begin{figure} [htb]
\centering
\begin{minipage}{0.496\linewidth}
\includegraphics[width=1.\linewidth]{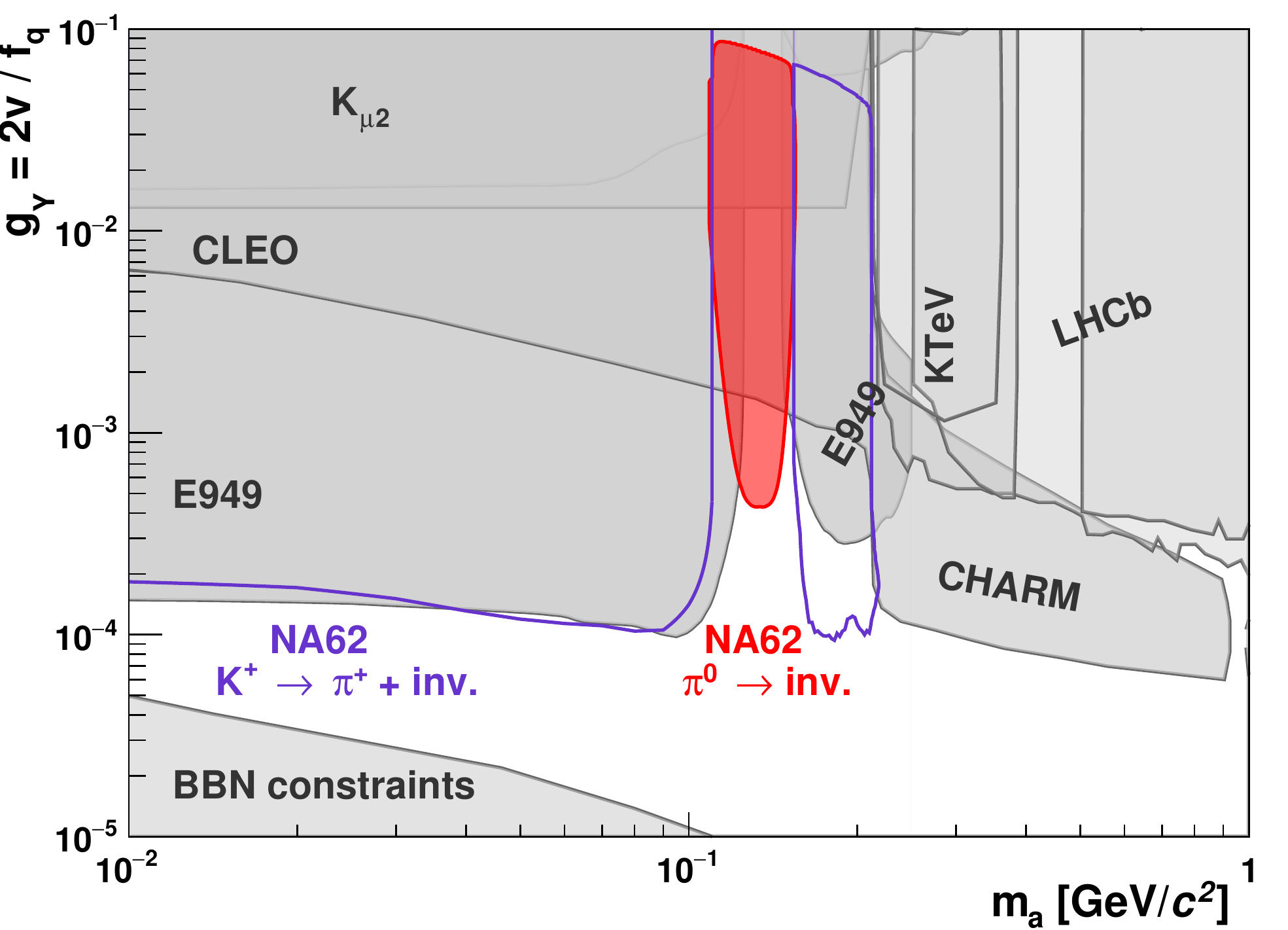}
\end{minipage}
\begin{minipage}{0.496\linewidth}
\includegraphics[width=1.\linewidth]{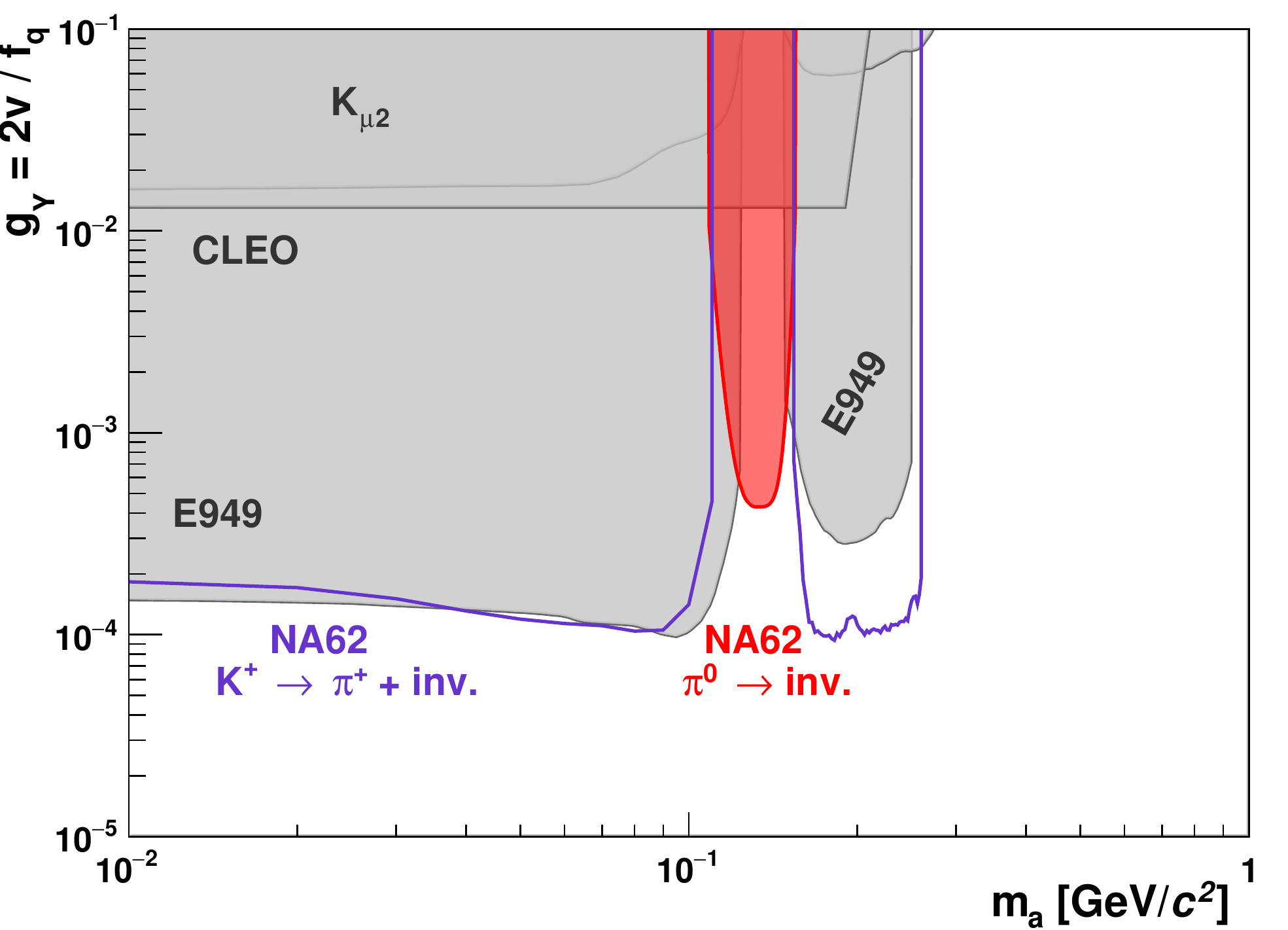}
\end{minipage}
\caption[]{Excluded regions of the parameter space ($m_a$, $g_Y$) for an ALP $a$ with SM interaction according to Equation~(\ref{eq:AxionPBCLagrangian}) dominantly decaying to SM particles (left, corresponding to the \textit{BC10} model of~\cite{RefPBC_Document}) and to invisible particles (right).
Bounds from the experiments E949~\cite{RefPBC_E949}, $K_{\mu 2}$~\cite{RefKm2Axion}, CLEO~\cite{RefCLEOAxion}, CHARM~\cite{RefPBC_CHARM}, KTeV~\cite{RefPBC_kTeV}, LHCb~\cite{RefPBC_LHCb_1,RefPBC_LHCb_2} and constraints from the Big Bang Nucleosynthesis are shown as grey areas.  
The exclusion bound from the present search for the decay $K^+ \rightarrow \pi^+ a$ in the $\pi^0$ mass region is labeled as ``NA62 $\pi^0 \rightarrow \mathrm{inv.}$''. The exclusion bound from the NA62 search for the decay $K^+ \rightarrow \pi^+ a$ outside the $\pi^0$ mass region is also shown~\cite{RefNA62Pnn2017,RefNA62Paper}.}
\label{fig:npPBCALP}
\end{figure}

If, within the accessible mass range, the ALP width is assumed to be dominated by decays to SM particles, the ALP rest lifetime must be taken into account when converting the analysis result into an upper limit for the coupling $g_Y$ as a function of the ALP mass $m_a$. The present result excludes a previously unexplored region of the parameter space, as shown in the left panel of Figure~\ref{fig:npPBCALP}.

If, within the accessible mass range, the ALP width is assumed to be dominated by decays to invisible particles, the corresponding ALP lifetime can be assumed to be significantly lower than that of the previous scenario. Nevertheless, the branching fractions for ALP decays to SM particles would be suppressed, so that the ALP would effectively be an invisible particle for the whole parameter space. In this scenario an even larger region of unexplored parameter space can be excluded, as shown in the right panel of Figure~\ref{fig:npPBCALP}: all the constraints from previous searches for ALP visible decays would be significantly weakened.

In~\cite{RefAxion_DMTaste}, different scenarios are considered for the ALP couplings to SM fermions. The exclusion limit from the present result can be seen to extend to an unexplored region of the parameter space for quark-universal models, when the ALP couples universally and exclusively to all SM quarks. However, the present result is already excluded by constraints from $B$ physics in quark-third-generation models, when the ALP couples universally and exclusively to bottom and top quarks.

It should be pointed out that the introduction of a light, feebly-coupled, spin-1 boson $U$ beyond the SM can effectively generate through its axial couplings the phenomenology related to an invisible spin-0 axion-like particle. This holds both in the context of supersymmetry, where the $U$ boson might be a spin-1 partner of the goldstino~\cite{Fayet1981}, or in a generalized scenario.
%
\subsection{Interpretation in terms of dark scalar production from \boldmath{$K^+$} decays}
\begin{figure} [htb]
\centering
\begin{minipage}{0.496\linewidth}
\includegraphics[width=1.\linewidth]{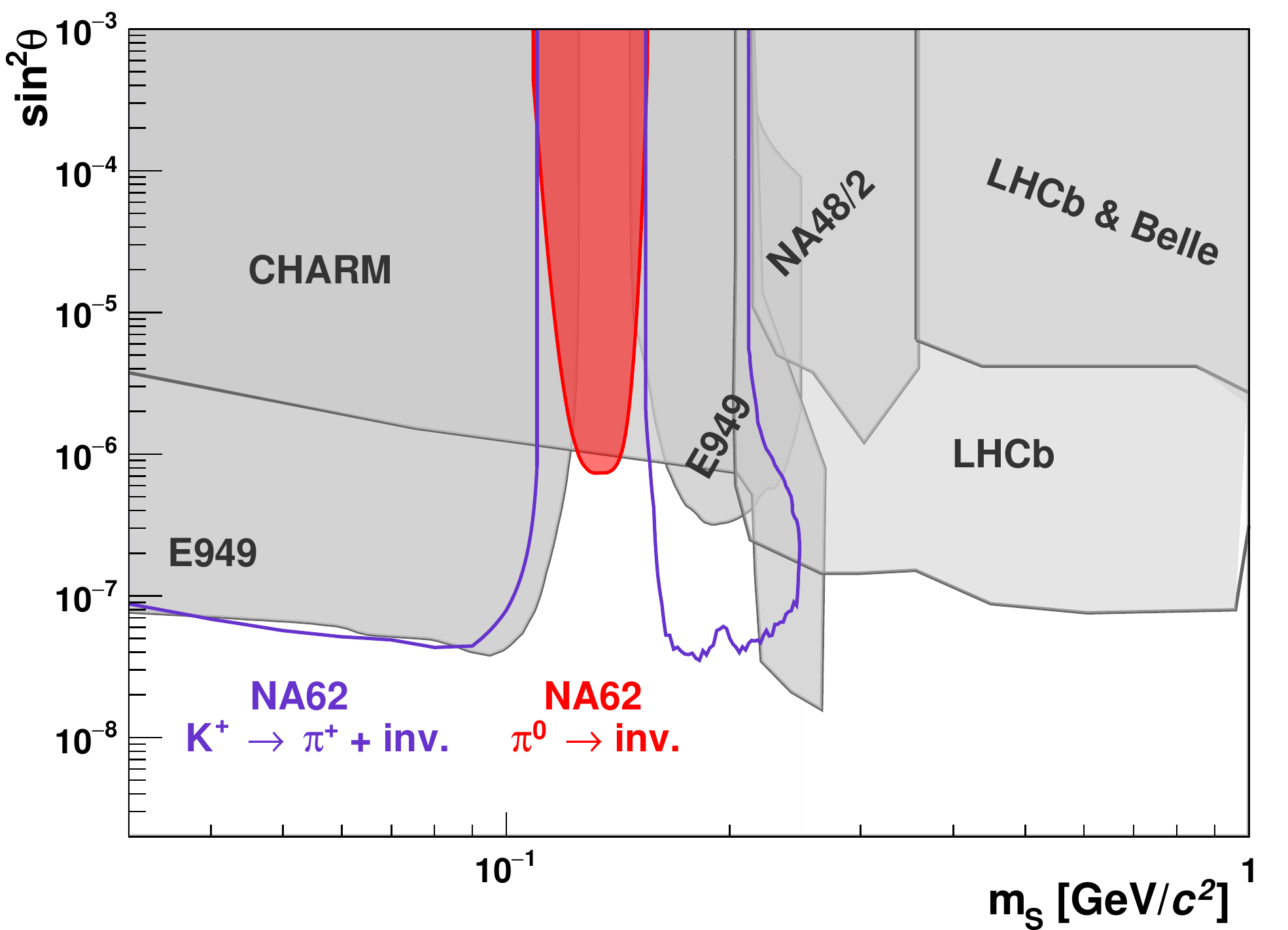}
\end{minipage}
\begin{minipage}{0.496\linewidth}
\includegraphics[width=1.\linewidth]{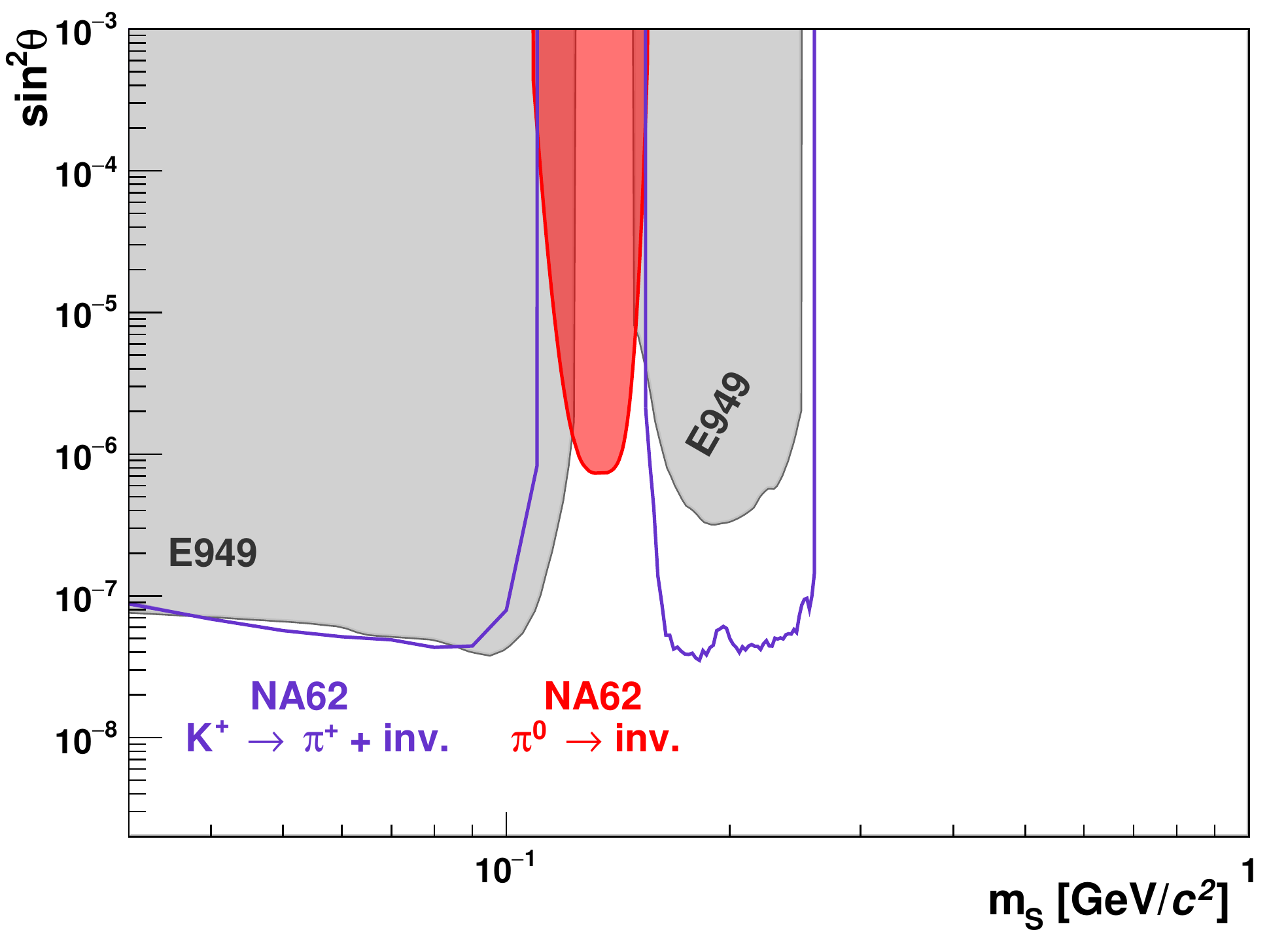}
\end{minipage}
\caption[]{
Excluded regions of the parameter space ($m_S$, $\sin^2\theta$) for a dark scalar $S$ with SM interaction according to Equation~(\ref{eq:ScalarLagrangian}) dominantly decaying to SM particles (left, corresponding to the \textit{BC4} model of~\cite{RefPBC_Document}) and to invisible particles (right).
Bounds from the experiments E949~\cite{RefPBC_E949}, CHARM~\cite{RefPBC_CHARM_1}, NA48/2~\cite{RefPBC_NA482}, LHCb~\cite{RefPBC_LHCb_1,RefPBC_LHCb_2}, Belle~\cite{RefPBC_Belle} 
are shown as grey areas.
The exclusion bound from the present search for the decay $K^+ \rightarrow \pi^+ S$ in the $\pi^0$ mass region is labeled as ``NA62 $\pi^0 \rightarrow \mathrm{inv.}$''. The exclusion bound from the NA62 search for the decay $K^+ \rightarrow \pi^+ S$ outside the $\pi^0$ mass region is also shown~\cite{RefNA62Pnn2017,RefNA62Paper}.
}
\label{fig:npPBCScalar}
\end{figure}

Following~\cite{RefPBC_Document}, the production of light dark scalars $S$ is investigated in the channel $K^+ \rightarrow \pi^+ S$. In a minimal model, one singlet field $S$ is considered,
\begin{equation}
\mathcal{L}_\mathrm{scalar} = - \left( \mu S + \lambda S^{2} \right) H^{\dagger}H,
\label{eq:ScalarLagrangian}
\end{equation}
where the coupling $\mu = \sin\theta$ relates to the $S$ mixing with the Higgs field $H$ and the $\lambda$ coupling to the interaction of the Higgs boson with a pair of scalars. The assumption $\lambda = 0$ is denoted as the \textit{BC4} scenario in~\cite{RefPBC_Document}, so that all production and decay processes of the dark scalar are controlled by the same parameter. The BR for $K^+\to \pi^{+}S$ in the \textit{BC4} model is evaluated according to~\cite{RefPBC_CHARM_1}.
If, within the accessible mass range, the width of the scalar is assumed to be dominated by decays to visible particles, a lifetime of $\simeq0.03\text{~ns}/\sin^2\theta$ is obtained. Within the range of $\sin^2\theta$ values shown in the left panel of Figure~\ref{fig:npPBCScalar}, infinite lifetimes for the dark scalar can safely be assumed. The result obtained extends into a region unexplored by previous experiments. If, as for the ALP search, dark scalar decays to invisible particles are assumed to dominate, the analysis results exclude an even larger region of unexplored parameter space, as shown in the right panel of Figure~\ref{fig:npPBCScalar}.

\section{Conclusions}
\label{sec:conclusions}
The hermetic, high-efficiency photon-veto system of the NA62 experiment has enabled the search for invisible decays of $\pi^0$ mesons tagged via the decay chain $\kppg$, $\dec$.
The veto inefficiency for background from $\pi^0$ decays is $3\times10^{-9}$ and no signal is found from the analysis of a sample of $4\times10^9$ tagged $\pi^0$ mesons.
The resulting 90\% CL upper limit on the branching ratio for $\dec$,
\begin{equation}
\mathrm{BR}(\dec)<4.4\times10^{-9},
\end{equation}
improves on previous results by a factor of 60. A model-independent limit is derived for the branching ratio of the decay $K^+\to\pi^+X$, where $X$ stands for any system with mass $m_X$ in the range 0.110--0.155~GeV/$c^2$ that is assumed to escape detection. 
The result is interpreted as a search for an axion-like particle (ALP, $a$) and a dark scalar ($S$) produced in the decay channels $K^+ \rightarrow \pi^+ a$ and $K^+ \rightarrow \pi^+ S$, respectively.
\clearpage 
\section*{Acknowledgements}
It is a pleasure to express our appreciation to the staff of the CERN laboratory and the technical
staff of the participating laboratories and universities for their efforts in the operation of the
experiment and data processing.

The cost of the experiment and its auxiliary systems was supported by the funding agencies of 
the Collaboration Institutes. We are particularly indebted to: 
F.R.S.-FNRS (Fonds de la Recherche Scientifique - FNRS), Belgium;
BMES (Ministry of Education, Youth and Science), Bulgaria;
NSERC (Natural Sciences and Engineering Research Council), funding SAPPJ-2018-0017 Canada;
NRC (National Research Council) contribution to TRIUMF, Canada;
MEYS (Ministry of Education, Youth and Sports),  Czech Republic;
BMBF (Bundesministerium f\"{u}r Bildung und Forschung) contracts 05H12UM5, 05H15UMCNA and 05H18UMCNA, Germany;
INFN  (Istituto Nazionale di Fisica Nucleare),  Italy;
MIUR (Ministero dell'Istruzione, dell'Univer\-sit\`a e della Ricerca),  Italy;
CONACyT  (Consejo Nacional de Ciencia y Tecnolog\'{i}a),  Mexico;
IFA (Institute of Atomic Physics) Romanian CERN-RO No.1/16.03.2016 and Nucleus Programme PN 19 06 01 04,  Romania;
INR-RAS (Institute for Nuclear Research of the Russian Academy of Sciences), Moscow, Russia; 
JINR (Joint Institute for Nuclear Research), Dubna, Russia; 
NRC (National Research Center)  ``Kurchatov Institute'' and MESRF (Ministry of Education and Science of the Russian Federation), Russia; 
MESRS  (Ministry of Education, Science, Research and Sport), Slovakia; 
CERN (European Organization for Nuclear Research), Switzerland; 
STFC (Science and Technology Facilities Council), United Kingdom;
NSF (National Science Foundation) Award Numbers 1506088 and 1806430,  U.S.A.;
ERC (European Research Council)  ``UniversaLepto'' advanced grant 268062, ``KaonLepton'' starting grant 336581, Europe.

Individuals have received support from:
Charles University Research Center (UNCE/SCI/ 013), Czech Republic;
Ministry of Education, Universities and Research (MIUR  ``Futuro in ricerca 2012''  grant RBFR12JF2Z, Project GAP), Italy;
Russian Foundation for Basic Research  (RFBR grants 18-32-00072, 18-32-00245), Russia; 
Russian Science Foundation (RSF 19-72-10096), Russia;
the Royal Society  (grants UF100308, UF0758946), United Kingdom;
STFC (Rutherford fellowships ST/J00412X/1, ST/M005798/1), United Kingdom;
ERC (grants 268062,  336581 and  starting grant 802836 ``AxScale'');
EU Horizon 2020 (Marie Sk\l{}odowska-Curie grants 701386, 842407, 893101).

%
\printbibliography[heading=bibintoc]
\end{document}